\documentclass[preprint]{aastex6} 
\newcommand{\diff}{\mathrm{d}}
\begin{document}

\title{Imaging Polarimeter for a sub-MeV Gamma-Ray all-sky survey \\using an Electron-Tracking Compton camera}

\author{
S.~Komura\altaffilmark{1}, A.~Takada\altaffilmark{1}, Y.~Mizumura\altaffilmark{2,1}, S.~Miyamoto\altaffilmark{1}, T.~Takemura\altaffilmark{1}, T.~Kishimoto\altaffilmark{1}, H.~Kubo\altaffilmark{1}, S.~Kurosawa\altaffilmark{3}, Y.~Matsuoka\altaffilmark{1}, K.~Miuchi\altaffilmark{4}, T.~Mizumoto\altaffilmark{1}, Y.~Nakamasu\altaffilmark{1}, K.~Nakamura\altaffilmark{1}, M.~Oda\altaffilmark{1},  J.~D.~Parker\altaffilmark{1}, T.~Sawano\altaffilmark{5}, S.~Sonoda\altaffilmark{1}, T.~Tanimori\altaffilmark{1,2}, D.~Tomono\altaffilmark{1}, and K.~Yoshikawa\altaffilmark{1}
}
\email{komura@cr.scphys.kyoto-u.ac.jp}

\altaffiltext{1}{Graduate School of Science, Kyoto University, Sakyo, Kyoto 606-8502, Japan}
\altaffiltext{2}{Unit of Synergetic Studies for Space, Kyoto University, Sakyo, Kyoto 606-8502, Japan}
\altaffiltext{3}{New Industry Creation Hatchery Center (NICHe), Tohoku University, Sendai, Miyagi, 980-8579, Japan}
\altaffiltext{4}{Department of Physics, Kobe University, Kobe, Hyogo, 658-8501, Japan}
\altaffiltext{5}{College of Science and Engineering, School of Mathematics and Physics, Kanazawa University, Kanazawa, Ishikawa, 920-1192, Japan
}

\begin{abstract}
X-ray and gamma-ray polarimetry is a promising tool to study the geometry and the magnetic configuration of various celestial objects, such as binary black holes or gamma-ray bursts (GRBs). 
However, statistically significant polarizations have been detected in few of the brightest objects. 
Even though future polarimeters using X-ray telescopes are expected to observe weak persistent sources, there are no effective approaches to survey transient and serendipitous sources with a wide field of view (FoV). 
Here we present an electron-tracking Compton camera (ETCC) as a highly-sensitive gamma-ray imaging polarimeter. 
The ETCC provides powerful background rejection and a high modulation factor over a FoV of up to 2$\pi$ sr thanks to its excellent imaging based on a well-defined point spread function. 
Importantly, we demonstrated for the first time the stability of the modulation factor under realistic conditions of off-axis incidence and huge backgrounds using the SPring-8 polarized X-ray beam. 
The measured modulation factor of the ETCC was 0.65 $\pm$ 0.01 at 150 keV for an off-axis incidence with an oblique angle of 30$^\circ$ and was not degraded compared to the 0.58 $\pm$ 0.02 at 130 keV for on-axis incidence.
These measured results are consistent with the simulation results. 
Consequently, we found that the satellite-ETCC proposed in \cite{2015ApJ...810...28T} would provide all-sky surveys of weak persistent sources of 13 mCrab with 10\% polarization for a 10$^{7}$~s exposure and over 20 GRBs down to a $6\times10^{-6}$ erg cm$^{-2}$ fluence and 10\% polarization during a one-year observation.

\end{abstract}


\section{Introduction} \label{sec:introduction}
X-ray and gamma-ray polarimetry in astronomy is widely viewed as a new probe for important open questions about high-energy sources such as gamma-ray bursts (GRBs), binary black holes (BBHs), active galactic nuclei (AGNs), and pulsars.
For example, statistical observations of GRB polarizations in the energy range of several tens of keV to a few MeV will be able to constrain competitive emission models with different magnetic field structures, which current photometric and spectroscopic observations have difficulty constraining \citep{2009ApJ...698.1042T}.
In addition, accreting black hole (BH) systems including BBHs and AGNs are thought to emit linearly polarized X-rays and gamma-rays due to scattering processes in their accretion disks, and therefore the measurement of these polarization properties and energies will enable us to determine the corona geometry, which is too small to be spatially resolved by current imaging observations \citep{2010ApJ...712..908S}.
More examples are discussed in \cite{1997SSRv...82..309L}, \cite{2006ChJAS...6a.237M}, and \cite{2011APh....34..550K}.

Despite their scientific importance, statistically significant polarization results have been detected in only a few of the brightest celestial X-ray and gamma-ray objects over the past four decades. 
In the 1970s, the Bragg-reflection X-ray polarimeter on board the {\it OSO-8} satellite first detected the polarization of the Crab nebula at 2.6 and 5.2 keV \citep{1978ApJ...220L.117W} and measured the upper limits for several X-ray objects \citep{1980ApJ...238..710L, 1984ApJ...280..255H}.
In the 2000s, two coded-mask detectors on board the {\it International Gamma-Ray Astrophysics Laboratory} ({\it INTEGRAL}) reported the polarization of the Crab nebula in the energy band between 0.1 MeV and 1 MeV \citep{2008Sci...321.1183D, 2008ApJ...688L..29F}; however, these results are plagued by large uncertainties because the instruments were not designed or calibrated for polarimetric observations. 
The {\it INTEGRAL} group also reported the energy dependence of the polarization fraction in Cygnus X-1; an upper limit of 20\% in the 250--400 keV band and a high polarization fraction of 67\% $\pm$ 30\% in the 0.4--2 MeV band were reported \citep{2011Sci...332..438L}. 
As for transient objects, several recent studies have reported that the prompt gamma-ray emission of several GRBs showed a high degree of polarization of 30--80\% and a time variation in the polarization direction \citep{2003Natur.423..415C, 2007ApJS..169...75K, 2007A&A...466..895M, 2009A&A...499..465M, 2009ApJ...695L.208G, 2013MNRAS.431.3550G, 2014MNRAS.444.2776G, 2011ApJ...743L..30Y, 2012ApJ...758L...1Y, 2016arXiv160807388R} in the energy band between 70 keV and 2 MeV. 
However, the statistical significances of these studies were marginal, typically having a confidence level of 2--3$\sigma$.

The current approaches to X-ray and gamma-ray polarimetry are classified roughly into two types.
The first is a pointing polarimeter that aims to observe persistent sources with a flux of 10--100 mCrab with high sensitivity.
In the energy range below $\sim$50 keV, both photoelectric and Compton polarimeters combined with X-ray focusing mirrors have been studied \citep{Weisskopf20161179, 2013ExA....36..523S, 2016NIMPA.838...89I, 2016APh....75....8K}.
X-ray mirrors can collect photons to a small detection area and dramatically reduce the background which causes serious degradation in the polarization sensitivity. 
In the energy range above $\sim$50 keV, Compton polarimeters with an active shield and a fine collimator to suppress the background have been studied \citep{2016APh....82...99C, 2016NIMPA.840...51K}.
The second approach consists of wide field of view (FoV) polarimeters with large detection area \citep{2009NIMPA.600..424B, 2011PASJ...63..625Y, 2011ASTRA...7...43O, 2014SPIE.9144E..4JG}; these are dedicated to observations of bright transient objects, especially prompt emissions of typical GRBs which last a few seconds with a fluence of approximately 10$^{-5}$ erg cm$^{-2}$. 
Even though a wide FoV increases the chance of GRB detection, it also accepts a huge background contribution coming from all directions. 
Therefore, GRB polarimeters have difficulty observing low signal-to-noise ratio sources, such as persistent sources and long-duration GRBs which last several tens of seconds or more.
In addition, owing to the lack of imaging capabilities, they essentially rely on other satellites to know the direction of the target sources, which would reduce the number of GRBs to be measured.
Even though the localization technique for bright GRBs by the relative count rates in instruments has been studied \citep{2010NIMPA.624..624S}, its accuracy would highly depend on the statistics of counts and background condition, and it can not be applied to other astronomical sources of course.
As seen above, there are no promising approaches to simultaneously explore both persistent and transient polarized sources in the universe.
An X-ray or gamma-ray polarimeter with both a moderate sensitivity and a wide FoV is required.

In the energy range from a few hundreds of keV to a few tens of MeV, Compton cameras have been studied as gamma-ray imaging spectroscopic telescopes capable of polarimetry and wide FoV.
A clear gamma-ray image based on a well-defined point spread function (PSF) can provide powerful background suppression by constraining the direction of incident photons. 
However, the Imaging Compton telescope (COMTEPL) \citep{1993ApJS...86..657S}, the only satellite-borne Compton camera, eventually indicated that it is difficult to reduce the background sufficiently using gamma-ray images obtained via conventional Compton cameras \citep{2001A&A...368..347W, 2004NewAR..48..193S}.
One approach to obtain a better quality of image than COMPTEL is improving the spatial and energy resolution of detectors \citep{2004NewAR..48..193S}.
This idea underlies the Compton Spectrometer and Imager (COSI), which is the Ge-based Compton camera designed to study nuclear-line emission and polarization \citep{2004NewAR..48..251B, 2015NIMPA.784..359C}.
Very recently, they succeeded to obtain the gamma-ray images of a few celestial objects and transients including one GRB in the long-duration balloon experiment, and they mentioned that the polarization analysis of those sources is underway \citep{2017arXiv170105558K}.
Another promised approach to improve the gamma-ray image of Compton camera is measuring the initial direction of the Compton recoil electron \citep{2004NewAR..48..193S}.
Many groups have proposed and studied the Compton camera with an electron tracker using the stacked solid-state detectors, which is designed to measure the recoil electron with an energy of more than a few MeV \citep{1996A&AS..120C.661O, 2002NewAR..46..611B, 2004NewAR..48..293K, 2015arXiv150807349M, 2016NIMPA.835...74K, 2016SPIE.9905E..2NT}.
On these types of Compton cameras, only the Medium Energy Gamma-Ray Astronomy telescope (MEGA) succeeded the demonstration of gamma-ray imaging polarimetry for on-axis incidence of 100\% polarized pencil beams at different energies (0.7, 2, and 5 MeV) \citep{2004ESASP.552..921Z}, where the beam images were reconstructed without (at 0.7 and 2 MeV) and with (at 5 MeV) electron tracks \citep{2005ExA....20..395A}.
%1996A&AS..120C.661O, TIGRE
%2002NewAR..46..611B, MEGA
%2004NewAR..48..293K, ACT
%2015arXiv150807349M, ComPair
%2016NIMPA.835...74K, PACT
%2016SPIE.9905E..2NT, ASTROGAM

In contrast to the trend using solid-state electron trackers, we have demonstrated the performance of an electron-tracking Compton camera (ETCC) utilizing a gaseous three-dimensional electron tracker since 2004 \citep{2004NewAR..48..263T}.
The gaseous tracker enables us to reduce the multiple-scattering angles and to measure the initial direction of the recoil electron more accurately than solid-state trackers.
As pointed out in a similar Compton camera concept with a gaseous tracker \citep{2004NewAR..48..299B}, such fine electron tracking is expected to reduce the PSF dramatically and consequently greatly improve the detection and polarization sensitivity.
In \cite{2015ApJ...810...28T}, we experimentally demonstrated that our ETCC has the ability to form a well-defined PSF of several degrees in the energy range from 100 keV to a few MeV.
Such a sharp PSF reduces a huge background contribution coming from all directions by nearly 3 orders of magnitude without any heavy shield \citep{Tanimori2017}, compared to typical non-imaging gamma-ray detectors such as coded-mask detectors and GRB polarimeters mentioned above.
In addition, particle identification using the energy-loss rate (dE/dx) of charged particles interacting in the gaseous detector is also possible and allows the rejection of non-gamma-ray backgrounds including neutrons \citep{2014JInst...9C5045M}. 
The satellite model ETCC is expected to have an effective area of 240 cm$^2$ with a PSF of 2$^{\circ}$ at 1 MeV, and the detection sensitivity would reach 1 mCrab flux at 1 MeV in a 10$^{6}$ s observation \citep{2015ApJ...810...28T}. 
Thanks to its powerful background suppression and wide FoV of up to 2$\pi$ sr \citep{2015JInst..10C1053M}, an ETCC has the capabilities of a highly-sensitive gamma-ray polarimeter that can be used not only to survey new faint persistent sources but also to observe transient objects including GRBs. 
In this paper, we investigate the basic polarimetric performance of the ETCC using both Monte Carlo simulations and experiments performed in the linearly polarized hard X-ray beamline at SPring-8. 
In Section \ref{sec:etcc}, we describe the concept of the ETCC as a Compton polarimeter and the setup of the Monte Carlo simulation; in Section \ref{sec:beam}, we report the analysis and results of the beam test compared to the simulation data; finally, we discuss the polarization sensitivities for future all-sky surveys using balloons and satellites in Section \ref{sec:discussion}.

%\clearpage

\section{Basic principles of Compton polarimetry} \label{sec:principles}
\begin{figure}[t]
\centering
\includegraphics[width=1.\textwidth]{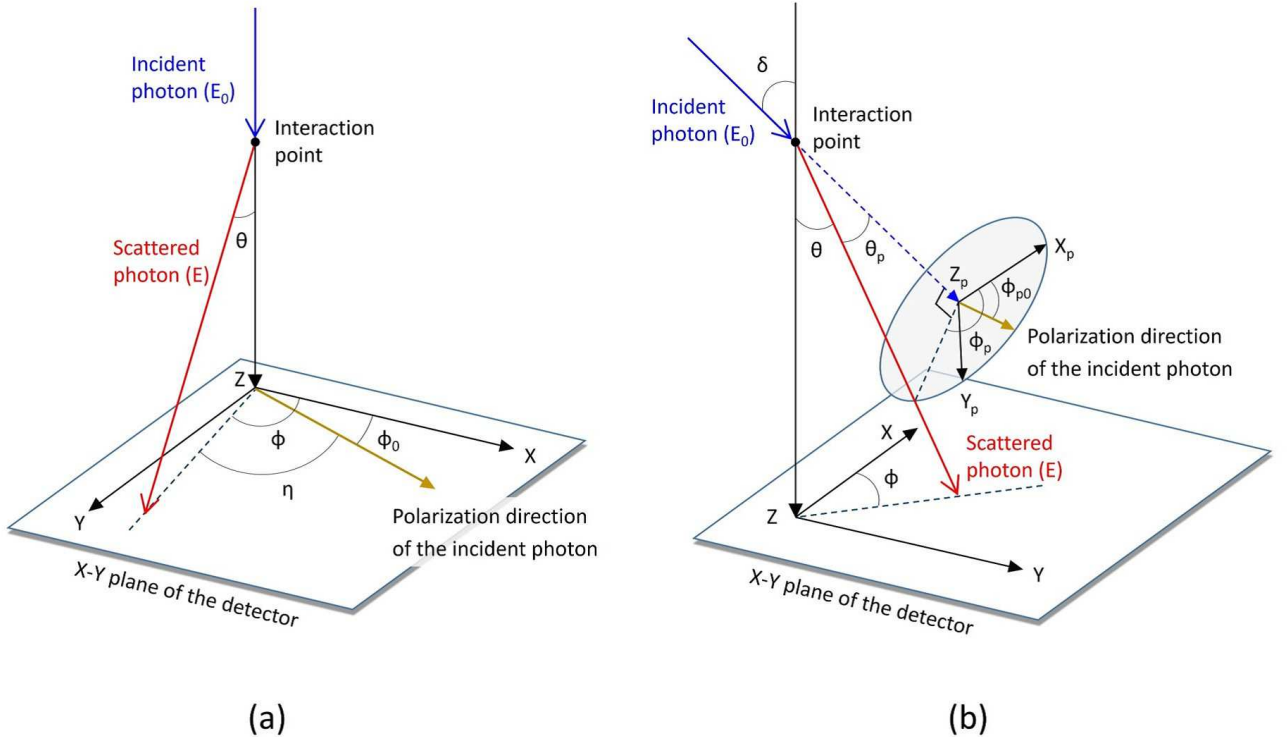}
\caption{Schematic of Compton scattering of a polarized photon. 
(a) When the incident direction is along the $Z$-axis of the detector coordinate system $XYZ$, the direction of the scattered photon is described using the polar angle $\theta$ and the azimuthal angle $\phi$.
(b) When the incident direction is not along the $Z$-axis, the direction of the scattered photon is described using the polar angle $\theta_{p}$ and the azimuthal angle $\phi_{p}$ in the photon coordinate system $X_{p}Y_{p}Z_{p}$.
The remaining symbols have the same meaning as in panel (a).
\label{fig:ComptonPolarimetry}
}
\end{figure}

In this section, we first show the principles necessary to measure polarization modulation with a Compton polarimeter and how it is affected by an off-axis incidence, mainly according to \cite{1997SSRv...82..309L}. Then, we reveal the requirements to maintain polarization sensitivity even for off-axis incidences.

Figure \ref{fig:ComptonPolarimetry}(a) shows a schematic view of Compton scattering of a polarized photon in which the incident direction of the photon is defined along the optical axis  ($Z$-axis) of the detector (in the case of on-axis incidence). 
In the detector coordinate system $XYZ$, which includes the $X$-$Y$ plane perpendicular to the $Z$-axis, the Compton scattering cross section for linearly polarized photons is expressed by the Klein-Nishina formula as
\begin{equation}
\frac{\diff \sigma}{\diff \Omega} = \frac{r_0^2}{2}\epsilon^2(\frac{1}{\epsilon}+\epsilon-2\sin^2\theta\cos^2\eta)
\label{eq:kleinnishina}
\end{equation}
where $r_0$ is the classical electron radius, $\epsilon$ is defined as $E_0/E$, $E_0$ and $E$ are the incident and scattered photon energies, respectively, $\theta$ is the polar angle of the scattered photon, and $\eta$ is the azimuthal angle of the scattered photon relative to the polarization direction of the incident photon \citep{1929ZPhy...52..853K}.  
According to this equation, photons dominantly scatter perpendicular to their polarization direction, and the angular distribution of the scattered photon $D(\theta, \eta)$ is strongly modulated.
By measuring this modulation, we can estimate the degree of polarization and the polarization direction of the incident photons.
In most cases, $\eta$ cannot be measured directly and is replaced by $\phi-\phi_{0}$, where $\phi$ is the azimuthal angle of the scattered photon relative to the $X$-axis and $\phi_{0}$ is the polarization angle of the incident photon (an unknown constant). 
Most Compton polarimeters simply measure the azimuthal angle distribution of the scattered photon $N(\phi)$ integrated for $\theta$, which is defined as follows:
\begin{equation}
N(\phi) \equiv \int D(\theta, \phi) \sin{\theta}\diff \theta.
\label{eq:Nphi}
\end{equation}
The range of integration in $\theta$ is limited by the geometrical structure of the detector.
After the integration, $N(\phi)$ can be theoretically expressed as a function of $\cos(2\phi)$ \citep{1997SSRv...82..309L}:
\begin{equation}
N(\phi) = a_0 \cos(2(\phi-\phi_{0}-\frac{\pi}{2}))+a_1,
\label{eq:modulationcurve}
\end{equation}
where $a_0$ and $a_1$ are the amplitude and average of the cosine curve, respectively. 
The polarimetric modulation factor is calculated as $\mu = a_0/a_1$, which is proportional to the degree of polarization.
Therefore, the degree of polarization of the incident photon is obtained from $\mu/\mu_{100}$, where $\mu_{100}$ is the modulation factor for 100\% linearly polarized incident photons and is often used as an instrumental analyzing power for polarization. 

When the incident photon has an incident angle of $\delta$ relative to the optical axis of the detector (in the case of off-axis incidence), $N(\phi)$ is affected by a complicated dependence not only on the polarization but also on the incident direction and energy, and therefore it no longer follows a $\cos(2\phi)$ curve \citep{1997SSRv...82..309L, 2014ApJ...782...28M}.
To obtain a distribution that follows the form of Equation (\ref{eq:modulationcurve}) for all incident angles, we need to move to the photon coordinate system, $X_{p}Y_{p}Z_{p}$, which includes the $Z_{p}$-axis along the incident direction and the $X_{p}$--$Y_{p}$ plane perpendicular to the $Z_{p}$-axis, as shown in Figure \ref{fig:ComptonPolarimetry}(b).
The displacement of the scattered photon in the $X_{p}Y_{p}Z_{p}$ system from the $XYZ$ system is calculated by 
the transformation matrix in the Cartesian coordinate system when the polar and azimuthal angles of the incident photon are known \citep{1997SSRv...82..309L, 2014ApJ...782...28M}.
Now, we can calculate the angular distribution of the scattered photon $D(\theta_{p}, \phi_{p})$ and the integrated azimuthal angle distribution $N(\phi_{p})$, where $\theta_{p}$ is the polar angle of the scattered photon and $\phi_{p}$ is the azimuthal angle of the scattered photon in the $X_{p}Y_{p}Z_{p}$ system. $N(\phi_{p})$ follows a $\cos(2\phi_{p})$ curve, and the modulation factor is calculated in the same form as mentioned above using the parameters of the curve.

Note that the $N(\phi)$ measured in any Compton polarimeter is distorted by systematic modulations due to the non-uniformity of the detector, and it is difficult to fit $N(\phi)$ with Equation~(\ref{eq:modulationcurve}).
Even for non-polarized incident photons, a fake modulation appears due to the non-uniformity of the detector response along the azimuthal angle $\phi$ (the effect of the non-uniform response).
In most cases, the systematic effect due to the off-axis incidence mentioned above occurs simultaneously (the effect of the off-axis incidence).
To cancel out these systematic modulations and obtain the corrected azimuthal angle distribution $N_{cor}$, we need to perform the following steps sequentially \citep{1997SSRv...82..309L}.
\begin{description}
\item[Step 1. Cancellation of the effect of off-axis incidence] \mbox{}\\ 
Calculate $N(\phi_{p})$ from the measured $D(\theta, \phi)$ using the coordinate transformation matrix.
\item[Step 2. Cancellation of the effect of non-uniform response] \mbox{}\\ 
Divide $N(\phi_{p})$ by the response for a non-polarized photon $N_{non}(\phi_{p})$:
\begin{equation}
N_{cor}(\phi_{p}) = \frac{N(\phi_{p})}{N_{non}(\phi_{p})}.
\label{eq:efficiency_correction}
\end{equation}
$N_{non}(\phi_{p})$ is the integrated azimuthal angle distribution in the $X_{p}Y_{p}Z_{p}$ system with a given energy and arrival direction that is equivalent to the ones observed; this needs to be calculated using a Monte Carlo simulation.
\end{description}

However, to avoid geometrical complexity, most Compton polarimeters do not measure the polar angle $\theta$.
Therefore, it is difficult to perform the coordinate transformation and cancel out the effect of off-axis incidence. 
The modulation factor naturally depends on the direction of the incident photons. 
In fact, several authors have reported that the modulation factor of their developed GRB polarimeters decreased by approximately 40\% for an incident angle of 60$^{\circ}$ in Monte Carlo simulations \citep{2009NIMPA.606..552X, 2009SPIE.7435E..0JM, 2014SPIE.9144E..4JG}.
These degradations in the modulation factor can be theoretically canceled out by the above two steps, and therefore we need a multipurpose polarimeter that measures all the required information, such as the three-dimensional direction of the scattered photons and the arrival direction and energy of the incident photon for each event.

If the modulation factor is obtained, the minimum detectable polarization (MDP) at the 99\% confidence level \citep{2010SPIE.7732E..0EW} can be calculated as follows and is commonly used as the polarimetric sensitivity of the detector:
\begin{equation}
MDP[\%] = \frac{429}{\mu_{100} R_{S}}\sqrt{\frac{R_{S}+R_{B}}{T}},
\label{eq:mdp}
\end{equation}
where $T$ is the exposure time of the observation, and $R_{S}$ and $R_{B}$ are the assumed count rate of the signal photons and backgrounds including both photon and non-photon particles, respectively. 
For observations of a point source, $R_{S}$ and $R_{B}$ correspond to $F_{S}A$ and $I_{B}\Delta\Omega A$, respectively, using the signal photon flux $F_{S}$ and the background intensity $I_{B}$. 
$A$ and $\Delta\Omega$ are the effective area and the angular resolution of the detector, respectively.
If the signal count rate is sufficiently larger than that of the background ($R_{S} \gg R_{B}$), the MDP improves in proportion to the modulation factor and the square root of the effective area ($MDP \propto 1/(\mu_{100} \sqrt A)$). 
Therefore, polarimeters designed to observe bright transient objects such as GRBs have large modulation factors and effective areas. Conversely, if we assume observations of persistent sources where the background is dominant ($R_{B} \gg R_{S}$), the MDP degrades in proportion to the square root of the angular resolution ($MDP \propto \sqrt {\Delta\Omega}/(\mu_{100} \sqrt A)$). 
Therefore, to measure the polarization of persistent sources, good angular resolution is also required.

%\clearpage

\section{Electron-Tracking Compton Camera as a polarimeter} \label{sec:etcc}
\begin{figure}[t]
\centering
\includegraphics[width=.6\textwidth]{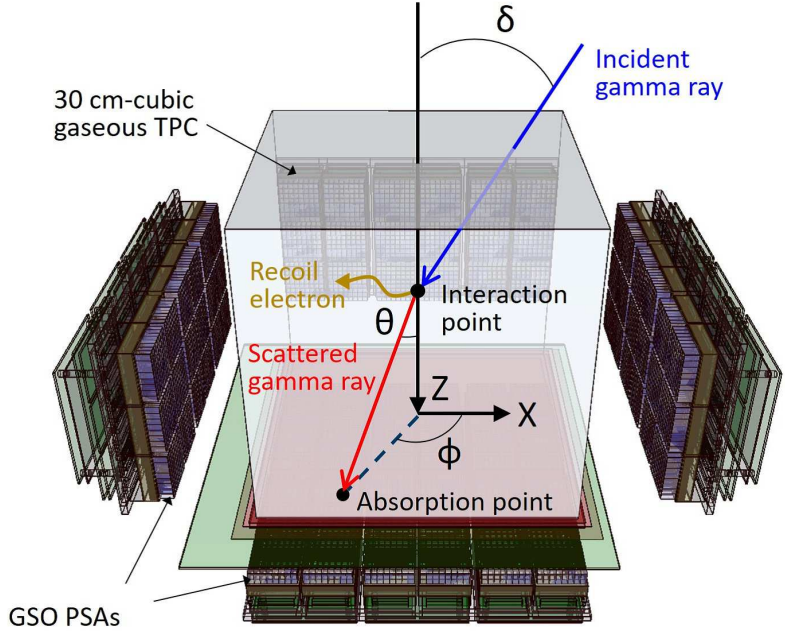}
\caption{Schematic view of the ETCC, which consists of a 30 cm-cubic gaseous time projection chamber (TPC) and pixel scintillator arrays (PSAs).
The TPC detects the track and energy of the recoil electron, and the PSAs detect the pixel position of the absorption and the energy of the scattered gamma-rays event by event. 
The $X$-$Y$ plane of the ETCC is defined to be parallel to the bottom plane of the TPC, and the $Z$-axis is defined to be perpendicular to the $X$-$Y$ plane. 
The Compton scattering angles $\theta$ and $\phi$, and the incident angle $\delta$ are also indicated as defined in Figure \ref{fig:ComptonPolarimetry}(b).
\label{fig:ETCC_SchematicView}}
\end{figure}

Standard Compton cameras were originally designed as versatile gamma-ray imagers that reconstruct the energy and arrival directions of incident gamma-rays based on Compton kinematics.
An ETCC is an advanced Compton camera that measures fine electron tracks using a gaseous time projection chamber (TPC) based on a micro-pattern gas detector. % which is set at the bottom of the TPC.
The measurement of fine electron tracks makes it possible to define the PSF with a small uncertainty based on a complete reconstruction of the Compton kinematics and to provide efficient background rejection based on the directional selection.
Figure \ref{fig:ETCC_SchematicView} shows a schematic view of the ETCC. Gd2SiO5:Ce (GSO) pixel scintillator arrays (PSAs), which act as absorbers for scattered gamma-rays, are set under the bottom and at each sides of the 30 cm-cubic TPC which is filled with an Ar-based gas. 
The ETCC already has sufficient sensitivity to measure the Crab signal at a 5$\sigma$ level in a 5-hour balloon observation \citep{2015ApJ...810...28T}.
Details of its design and performances are described in \cite{2014JInst...9C5045M}, \cite{2015JInst..10C1053M}, \cite{2015NIMPA.800...40M}, and \cite{2015ApJ...810...28T, Tanimori2017}. 
Based on the current analysis, the Compton interaction point in the TPC is determined by a track fitting analysis with a spatial resolution of less than 1 cm \citep{takadaDth}, and the absorption positions of scattered gamma-rays in the PSAs are determined within the size of the pixel scintillator (6$\times$6$\times$13 mm$^3$). 

In general, Compton cameras, including the ETCC, can be used as Compton polarimeters because the Compton camera determines the three-dimensional direction of the scattered gamma-ray as the direction from the interaction point to the absorption point and obtains the angular distribution of the scattered gamma-ray $D(\theta, \phi)$ for each gamma-ray.
In addition, the ETCC uniquely determines the incident direction event by event; therefore, $\theta_{p}$ and $\phi_{p}$ in the photon coordinate system can be geometrically calculated.
Therefore, the ETCC can correct for the effect of an off-axis incidence and has a large FoV for the polarization measurement.
Furthermore, due to powerful background suppressions with a sharp PSF and particle identification of the electron tracks, we expect a much better MDP than that of standard Compton cameras even in intense background conditions, such as space. 

To calculate the MDPs of the ETCC, we need to estimate the modulation factor and the detection efficiency for various energetic gamma-rays, incident directions, and polarization directions using Geant4 simulations \citep{2003NIMPA.506..250A} with a detailed geometrical model of the current ETCC \citep{2014JPSCP...1a3099S}. 
The performance of the simulation has already been checked with ground calibration results using non-polarized gamma-ray sources \citep{2015ApJ...810...28T}. 
In this study, we used the physics models in Geant4 called {\tt G4EmLivermorePolarizedPhysics} to account for polarized low energy gamma-rays.
This simulation provides the Compton interaction point in the TPC, the pixel position of the absorption in the PSAs for the scattered photons, and the energies of the scattered photon and recoil electron for each incident photon.
As it does not include the electron tracking in the TPC, we do not take account of the uncertainty of the interaction point, which is not serious because its effects on the $\theta_{p}$ and $\phi_{p}$ for each photon are small, typically 1$^{\circ}$ and a few degrees, respectively.
The energy resolutions of the TPC and PSAs are included in the simulations, assuming that they follow Gaussian distribution with the Full Width at Half Maximum (FWHM) of 22\% for 22 keV and 11\% for 662 keV, respectively.

\begin{figure}[t]
  \centering
  \sbox0{\includegraphics{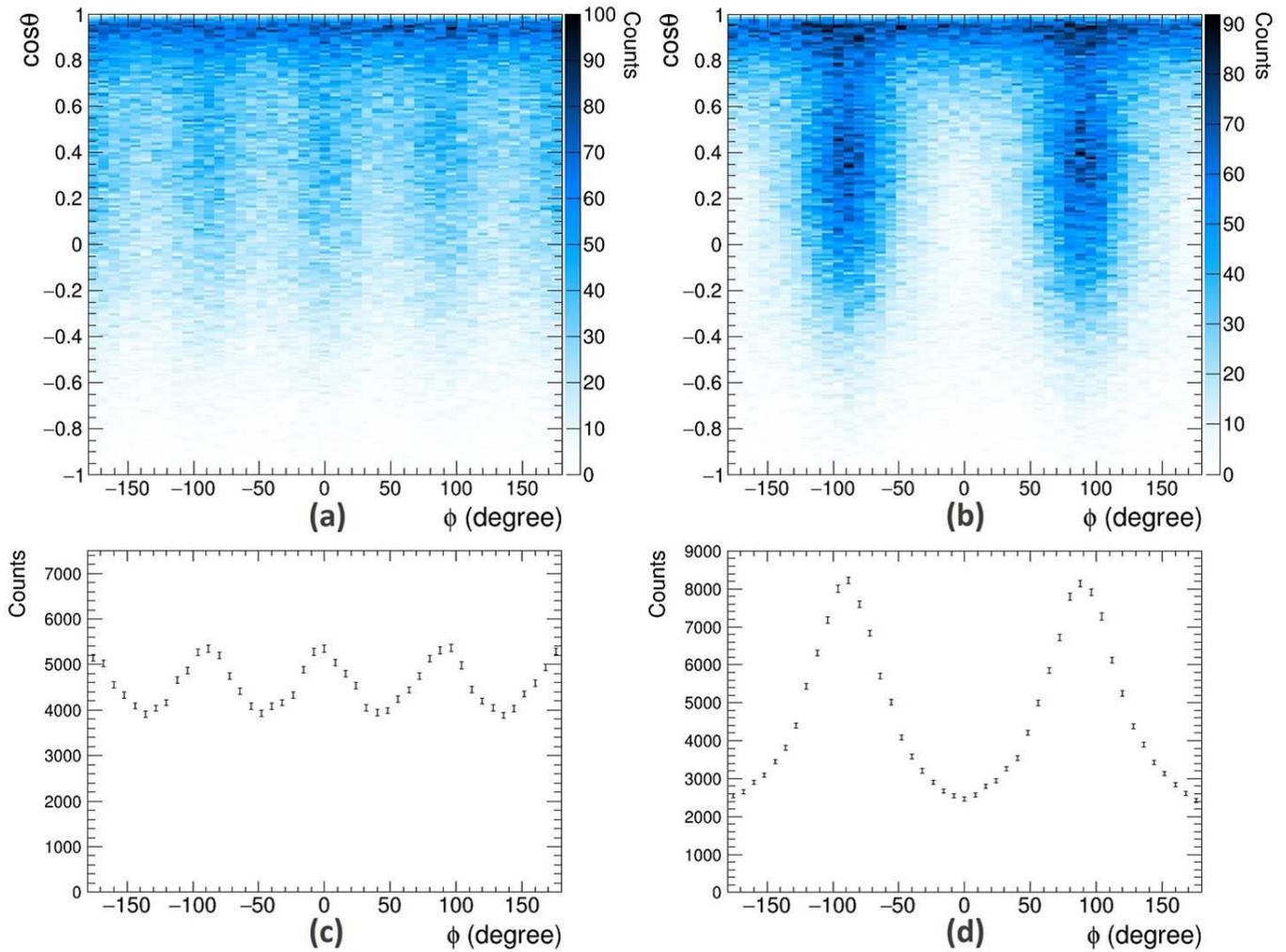}}
  \includegraphics[clip, trim={0 0 {.0\wd0} {.0\ht0}}, width=1.\textwidth]{SimulatedScatteredGammaDistributions.eps}
  \caption{
Simulated distributions of scattered gamma-rays for incident on-axis 200-keV gamma-rays for the ETCC. (a, b) Two-dimensional scatter plot of $\cos{\theta}$ and $\phi$ for non-polarized gamma-rays $D_{non}(\cos{\theta}, \phi)$ and for 100\% linearly polarized gamma-rays $D_{pol}(\cos{\theta}, \phi)$, respectively. (c, d) Azimuthal event distributions integrated for the cos$\theta$ of non-polarized gamma-rays $N_{non}(\phi)$ and the same for 100\% linearly polarized gamma-rays $N_{pol}(\phi)$, respectively. 
The error bars in panels (c) and (d) represent the 1$\sigma$ statistical error.
}
  \label{fig:SimulatedScatteredGammaDistributions}
\end{figure}

\begin{figure}
  \centering
  \sbox0{\includegraphics{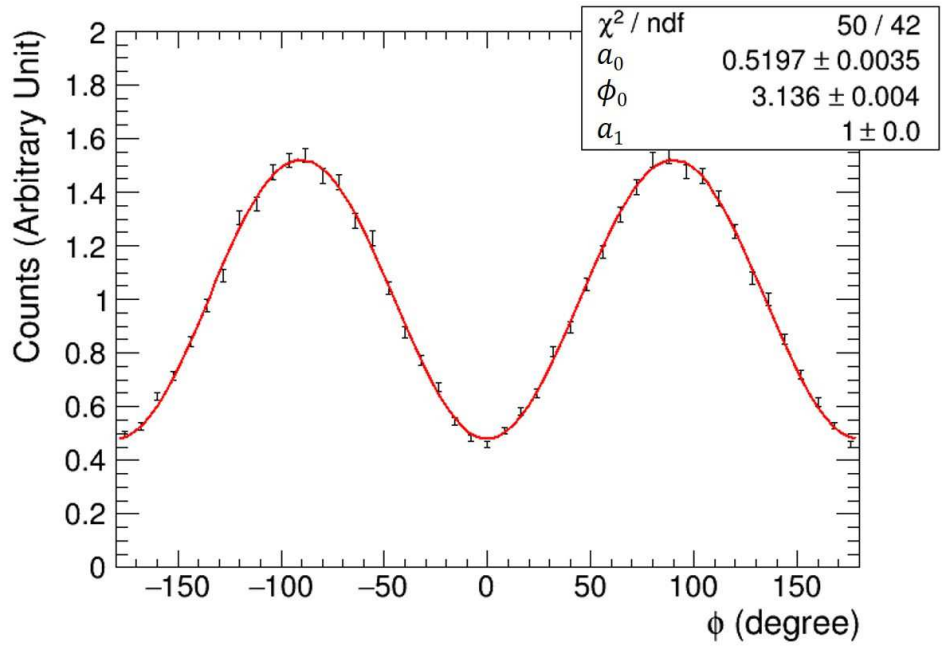}}
  \includegraphics[clip, trim={0 0 {.0\wd0} {.0\ht0}}, width=0.7\textwidth]{SimulatedModulationCurve_div.eps}
  \caption{
Corrected integrated azimuthal angle distribution of scattered photon $N_{cor}(\phi)$ calculated from $N_{pol}(\phi)/N_{non}(\phi)$.}
  \label{fig:SimulatedModulationCurve_div}
\end{figure}

After applying the event selection criteria described in \cite{2014JPSCP...1a3099S}, the angular distribution of scattered gamma-rays $D(\cos{\theta}, \phi)$ and the integrated azimuthal angle distribution $N(\phi)$ are obtained as shown in Figure \ref{fig:SimulatedScatteredGammaDistributions} for 200-keV incident gamma-rays, with an incident angle $\delta$ set to zero (on-axis incidence) and a polarization direction along the $X$-axis.
Figure \ref{fig:SimulatedScatteredGammaDistributions}(a) shows $D_{non}(\cos{\theta}, \phi)$ for non-polarized incident gamma-rays, and Figure \ref{fig:SimulatedScatteredGammaDistributions}(b) shows $D_{pol}(\cos{\theta}, \phi)$ for 100\% linearly polarized incident gamma-rays.
In both figures, Compton-scattered photons absorbed in the bottom PSAs are modulated near $\cos{\theta} = 1$ (i.e., forward scattering events).
In Figure \ref{fig:SimulatedScatteredGammaDistributions}(a), events absorbed in the four side PSAs are modulated near values of $\phi$ of $-$180$^\circ$, $-$90$^\circ$, 0$^\circ$, 90$^\circ$, and 180$^\circ$ in the range of $\cos{\theta}$ between $-$1.0 and 0.8.
Conversely, in Figure \ref{fig:SimulatedScatteredGammaDistributions}(b), multiple events are modulated near values of $\phi$ of $-$90$^\circ$ and 90$^\circ$, where the direction of the incident gamma-ray is perpendicular to the polarization direction of the incident photon and the Compton scattering cross section is at its maximum according to Equation (\ref{eq:kleinnishina}).
The azimuthal distributions $N(\phi)$s integrated over $\theta$ are calculated according to 
\begin{equation}
N(\phi) = \int_{-1}^{(\cos{\theta})_{max}} D(\cos{\theta}, \phi) \diff(\cos{\theta})
\label{eq:Nphi_cos}
\end{equation}
which is derived from Equation (\ref{eq:Nphi}). 
Figures \ref{fig:SimulatedScatteredGammaDistributions}(c) and \ref{fig:SimulatedScatteredGammaDistributions}(d) show $N_{non}(\phi)$ and $N_{pol}(\phi)$, respectively. 
%The calculated $N(\phi)$s from both panels (a) and (b) are shown in the left lower panel (c) and the right lower panel (d), respectively.
We found that, in $N_{non}(\phi)$, a small systematic modulation appears due to the non-uniformity of the detector response, as mentioned in Section \ref{sec:principles}. 
Figure \ref{fig:SimulatedModulationCurve_div} shows the azimuthal angle distribution $N_{cor}(\phi)$ corrected for the response effect according to Equation (\ref{eq:efficiency_correction}) with the best fit curve given by Equation (\ref{eq:modulationcurve}); the modulation factor of the ETCC is estimated to be 0.52 $\pm$ 0.01 for on-axis incident photons with energies of 200 keV.

\begin{figure}
  \centering
  \sbox0{\includegraphics{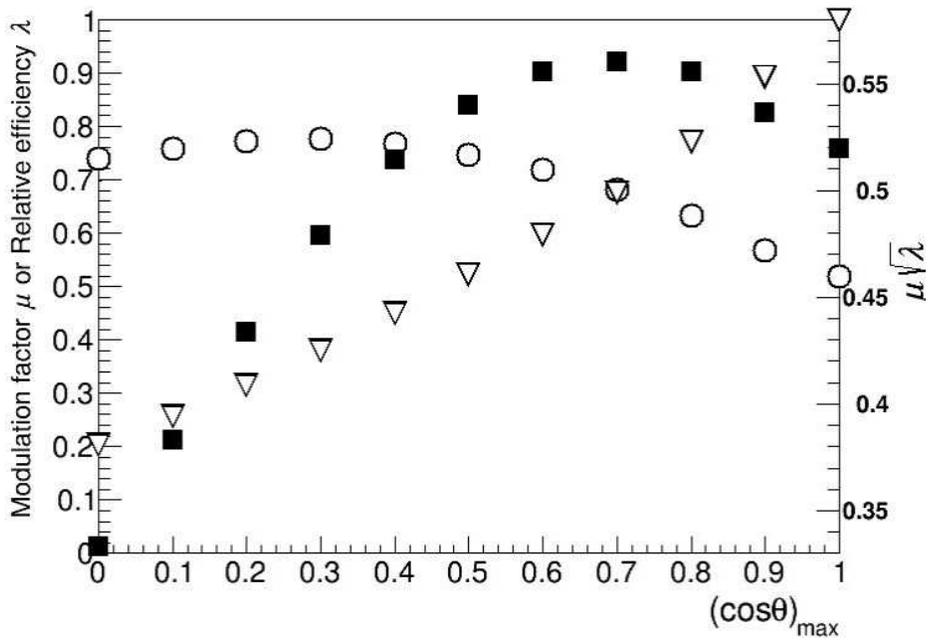}}
  \includegraphics[clip, trim={0 0 {.01\wd0} {.0\ht0}}, width=0.7\textwidth]{SimulatedModulationFactor_cos.eps}
\caption{Dependences of the modulation factor $\mu$ (open circles), the relative detection efficiency $\lambda$ (open triangles), and the figure of merits $\mu \sqrt{\lambda}$ (filled squares) on the integration region from 0 to $\theta_{max}$.
}
\label{fig:SimulatedModulationFactor_cos}
\end{figure}

As mentioned in Section \ref{sec:principles}, MDP is inversely proportional to $\mu_{100} \sqrt{A}$.
Figure \ref{fig:SimulatedModulationFactor_cos} shows the dependence of  $\mu$, the relative detection efficiency $\lambda$, and $\mu \sqrt{\lambda}$ on $(\cos{\theta})_{max}$ in Equation (\ref{eq:Nphi_cos}), 
where $\lambda$ is normalized to 1 at $(\cos{\theta})_{max} = 1$.
We found that a $(\cos{\theta})_{max}$ of 0.7 minimizes the MDP for on-axis incident photons with energies of 200 keV when $\mu \sqrt{\lambda}$ is at its maximum.
Of course, the optimal range of the integration also depends on the incident energy and incident angle, and therefore we need to minimize the MDP for each energy band.
For simplification in the following discussion, we calculate the integrated azimuthal distribution $N(\phi)$ in the range of $\cos{\theta}$ from $-$1.0 to 0.7.

\clearpage

\section{Experiments on a linearly polarized X-ray Beam}\label{sec:beam}
We performed two types of experiments from January 27--31, 2015, using the ETCC on the High Energy Inelastic Scattering Beamline BL08W at SPring-8, which supplies a $>$ 99\% linearly polarized hard X-ray beam with an energy of 182 keV.

\subsection{Polarization measurement for on-axis incidence} \label{sec:experiment1}
\begin{figure}[t]
  \centering
  \begingroup
  \sbox0{\includegraphics{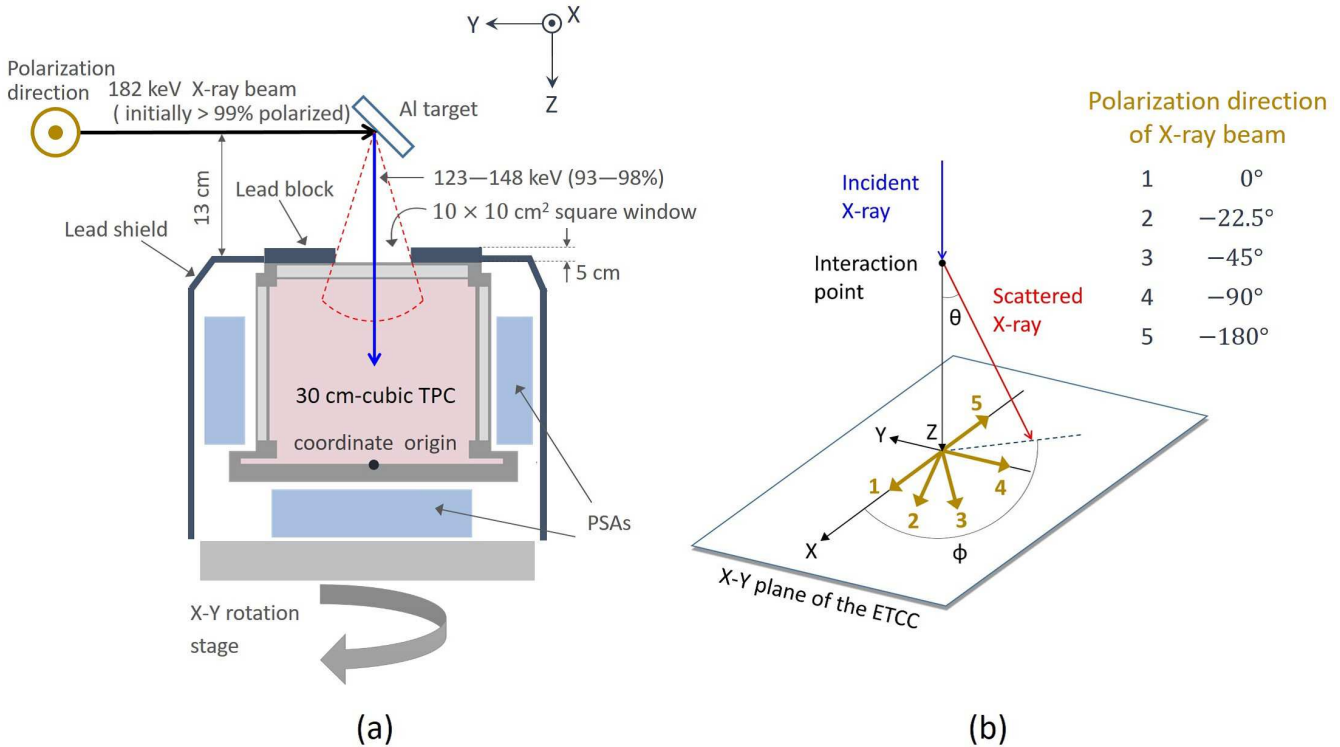}}
  \includegraphics[clip, trim={0 0 {.0\wd0} {.0\ht0}}, width=1.0\textwidth]{SPring-8Exp1Setup.eps}
  \endgroup
\caption{
(a) A side view in the $Y$-$Z$ plane of the current ETCC setup for the first experiments on BL08W at SPring-8. 
The origin of the coordinate system of the ETCC ($XYZ$) was set at the bottom center of the TPC.
(b) The ETCC measured the angular distribution $D(\theta, \phi)$ of the scattered X-ray in the TPC, where $\theta$ and $\phi$ are the polar scattering angle and the azimuthal scattering angle, respectively. 
The measurements were performed for five different polarization directions of the X-ray beam by rotating the ETCC in the $X$-$Y$ plane. 
The corresponding azimuthal angles are 0$^\circ$, $-$22.5$^\circ$, $-$45$^\circ$, $-$90$^\circ$, and $-$180$^\circ$.
}
\label{fig:SPring-8Exp1Setup}
\end{figure}

In the first experiment, we measured the modulation factor of the ETCC and compared it with our simulation results for an on-axis beam with various polarization directions.
The experimental setup is shown in Figure \ref{fig:SPring-8Exp1Setup}(a).
We irradiated the X-ray beam to a 10-mm thick aluminum (Al) target, from which X-rays scattered vertically at the target entered the ETCC. 
The (X, Y) coordinates of the beam spot on the Al target were set to (10 mm, 0 mm).
The intensity of the X-ray beam was considerably weakened by 20-cm thick Al attenuators set in front of the target at a distance of 137 cm. 
The front and back sides of the ETCC in the beam direction were shielded by 1-mm-thick lead sheets to reduce chance coincidence noise between the TPC and the PSAs due to ambient X-rays in the laboratory.
However, the side faces of the ETCC were not covered because of the lack of lead sheets; a large amount of chance coincidence noise occurred as described below.
Due to spatial limitations in the laboratory, the Al target was located at a height of 13 cm just above the ETCC. 
Scattered X-rays at the Al target were roughly collimated with an opening window (10 cm $\times$ 10 cm) in the lead blocks set at the top of the TPC, and therefore the energy and degree of polarization of the incident X-rays were widely spread from 123 to 148 keV and from 93\% to 98\%, respectively. 
As shown in Figure \ref{fig:SPring-8Exp1Setup}(b), the ETCC measured the angular distribution of the Compton scattered X-rays for five different incident X-ray polarization directions by rotating the ETCC around its $Z$-axis.
First, the azimuthal angle of the polarization direction of the X-ray beam was set to 0$^\circ$.
The event rate with the Al target (on-target measurement) was approximately 300 Hz and contained huge background levels due to air scattering, approximately three times larger than that expected at balloon altitude \citep{2015NIMPA.800...40M}.
We performed the measurement with no Al target (off-target measurement) for each polarization direction to subtract backgrounds from the on-target data in the off-line analysis.

\begin{figure}[]
  \centering
  \begingroup
  \sbox0{\includegraphics{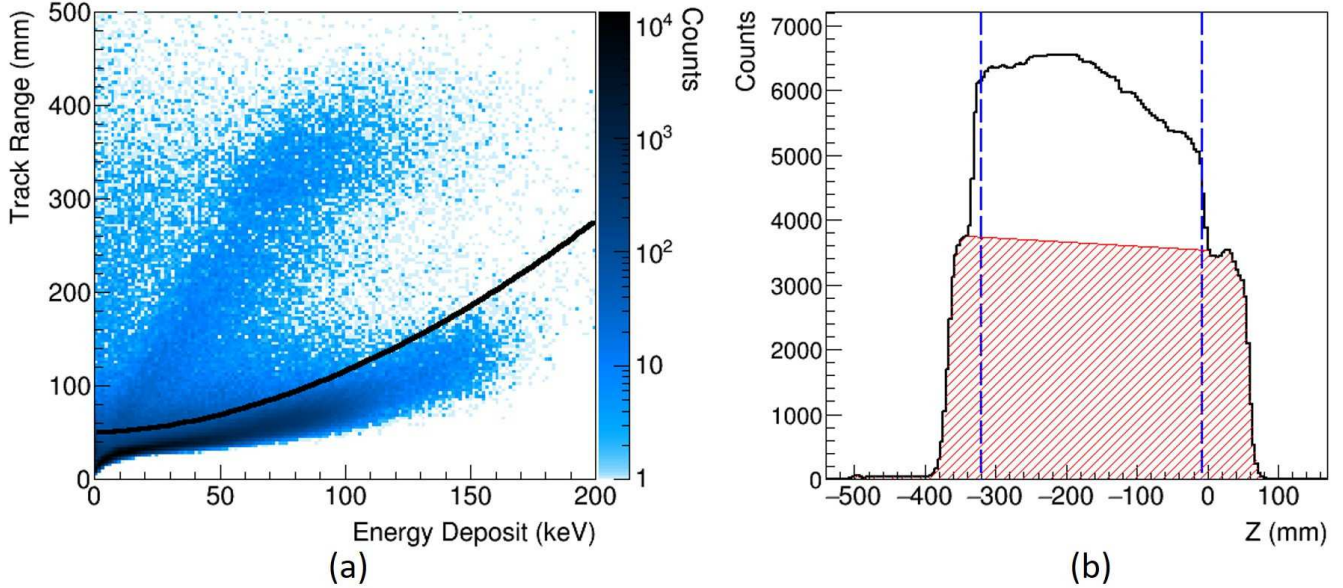}}
  \includegraphics[clip, trim={0 0 {.0\wd0} {0.0\ht0}}, width=1.\textwidth]{SPring-8Exp1EventSelection.eps}
  \endgroup
\caption{
(a) Two-dimensional plot of the measured track range and energy deposit in the TPC indicating the energy loss rate (dE/dx) of the charged particles.
The solid line represents the selection criteria described in Equation (\ref{eq:dedx}); the lower side of the line has fully contained electrons stopping in the TPC and the upper side has MIP-like charged particles, such as cosmic muons and high-energy electrons escaping from the TPC.
(b) Distribution of Compton scattering position in the TPC along the $Z$-axis direction (solid bold line). 
The hatched area is due to the chance coincidence noise. The two vertical dashed lines indicate the fiducial volume region used in this analysis.
\label{fig:SPring-8Exp1EventSelection}
}
\end{figure}

To obtain reconstructable Compton events, we performed the following event selections.
First, we selected the correct Compton event, where the recoil electron stops in the TPC, using the relationship between the measured track range and the energy deposited in the TPC \citep{2014JInst...9C5045M} as described below:
\begin{equation}
\left(\frac{{\rm Track\ Range}}{[{\rm mm}]}\right) < 4.1 \times 10^3 \left(\frac{{\rm Energy\ Deposit}}{[{\rm MeV}]}\right)^{1.8} + 50,
\label{eq:dedx}
\end{equation}
which is drawn as the solid line in Figure \ref{fig:SPring-8Exp1EventSelection}(a). 
Next, we selected the events interacting in the fiducial volume of the TPC. 
Figure \ref{fig:SPring-8Exp1EventSelection}(b) shows the distribution of the analyzed starting positions of the measured tracks (i.e., the Compton scattering positions) along the $Z$-axis.
The coincidence events between TPC and PSAs lay within the region of approximately $-$340 mm $< Z <$ 0 mm, which includes both the signal X-rays scattered on the inside of the TPC and the chance coincidence noise due to air scattering.
The remaining areas are all formed by the chance coincidence noise, in which the time lag between incidents on TPC and PSAs is longer than the time window of the coincidence.
We defined the fiducial gas volume region as $-$319.9 mm $< Z <$ $-$7.9 mm, and we selected events within this area.
Even though chance coincidence events remain after the above removal, we can still estimate the signal-to-noise ratio in the fiducial volume of the TPC by assuming that the chance coincidence events are approximately uniformly distributed along the $Z$-axis \citep{2015NIMPA.800...40M}, which is denoted as the hatched area in Figure \ref{fig:SPring-8Exp1EventSelection}(b).
In the case of Figure \ref{fig:SPring-8Exp1EventSelection}(b), we found that the selected events contain approximately 63\% of the noise, and the signal occupied only one third of the recorded data.
Therefore, the experiment was conducted under background dominant conditions.
Because the noise in the fiducial volume has the same features as the events lying outside the fiducial volume, we can subtract the noise component using these events.

\begin{figure}[]
  \centering
  \begingroup
  \sbox0{\includegraphics{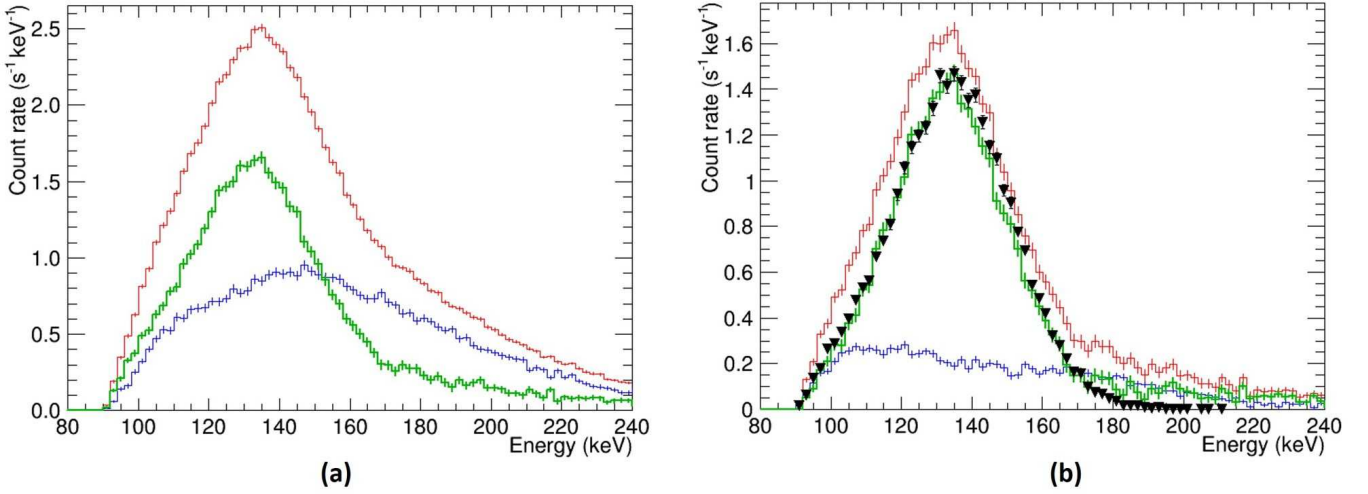}}
  \includegraphics[clip, trim={0 0 {.0\wd0} {0.\ht0}}, width=1.\textwidth]{SPring-8Exp1EnergySpectrum.eps}
  \endgroup
\caption{
(a) Subtraction of the chance coincidence noise from the measured energy spectrum of the incident X-rays. 
The energy spectrum of the Al on-target data after the fiducial volume selection (red), the chance coincidence noise (blue), and the residual events (green) after the subtraction of the chance coincidence noise are shown. The blue line is obtained by sampling and scaling the energy distribution of the events lying between $Z < -$340 mm and $Z >$ 0 mm.
(b) Subtraction of the Al off-target data and the final reconstructed energy spectrum.
The red and blue lines represent the Al on-target and off-target data, respectively, after the removal of the chance coincidence noise.
The green line represents the residual events after the subtraction of the off-target data, which is in good agreement with the simulated spectrum (filled triangles). 
The error bars in panels (a) and (b) represent the 1-sigma statistical error.
}
\label{fig:SPring-8Exp1EnergySpectrum}
\end{figure}

We confirmed the validity of the above selections using the measured energy spectra of the incident X-rays.
As shown in Figure \ref{fig:SPring-8Exp1EnergySpectrum}(a), the spectrum of Al on-target data after the fiducial volume selection expanded to a higher energy region than the expected energy range of 123--148 keV. 
This is because the chance coincidence noise includes air-scattered X-rays with relatively high energy, close to that of the X-ray beam energy, 182 keV.
After subtracting the chance coincidence noise, we found that the residual energy spectrum is concentrated near the expected energy range.
In addition, by subtracting the energy spectrum of off-target measurement, an energy peak near 130 keV appeared, as shown in Figure \ref{fig:SPring-8Exp1EnergySpectrum}(b), which is consistent with the expected energy of incident X-rays scattering at 90$^\circ$ from the Al target, 134 keV; in addition, there is good consistency between the measured and simulated energy spectrums within 10\% below 170 keV.

\begin{figure}[h]
  \centering
  \begingroup
  \sbox0{\includegraphics{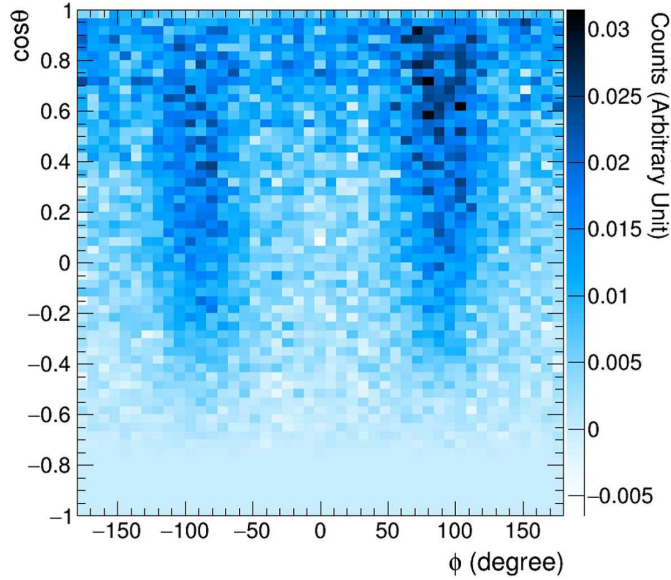}}
  \includegraphics[clip, trim={0 0 {.0\wd0} {.0\ht0}}, width=0.5\textwidth]{SPring-8Exp1ScatteredGammaDistribution.eps}
  \endgroup
\caption{
Two-dimensional scatter plot of the measured $D^{mes}_{pol}(\cos{\theta}, \phi)$ when the polarization direction of the X-ray beam is 0$^\circ$. 
\label{fig:SPring-8Exp1ScatteredGammaDistribution}}
\end{figure}

\begin{figure}[h]
  \centering
  \begingroup
  \sbox0{\includegraphics{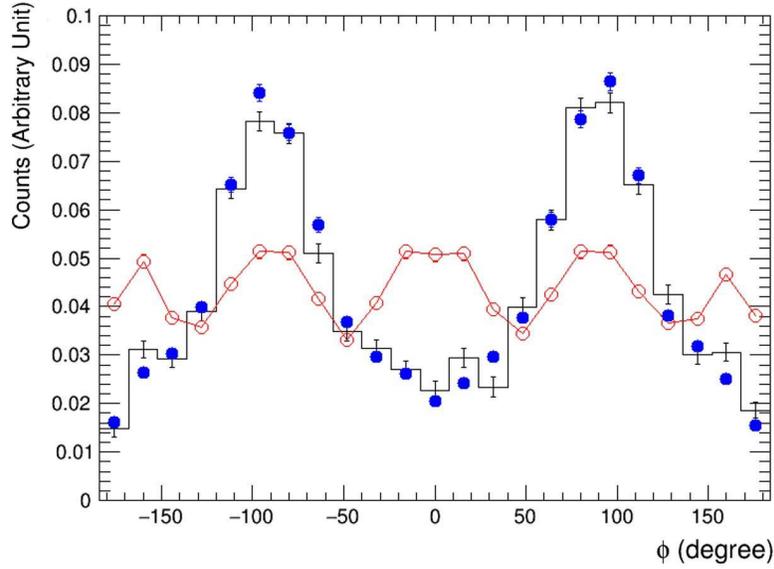}}
  \includegraphics[clip, trim={0 0 {.0\wd0} {.0\ht0}}, width=0.6\textwidth]{SPring-8Exp1AzimuthalDistribution.eps}
  \endgroup
\caption{
The solid line histogram represents the measured azimuthal angle distribution $N^{mes}_{pol}(\phi)$.
The simulation results of the azimuthal angle distribution $N^{sim}_{pol}(\phi)$ and $N^{sim}_{unpol}(\phi)$ are plotted as filled circles and open circles, respectively.
\label{fig:SPring-8Exp1AzimuthalDistribution}}
\end{figure}

\begin{figure}[h]
  \centering
  \begingroup
  \sbox0{\includegraphics{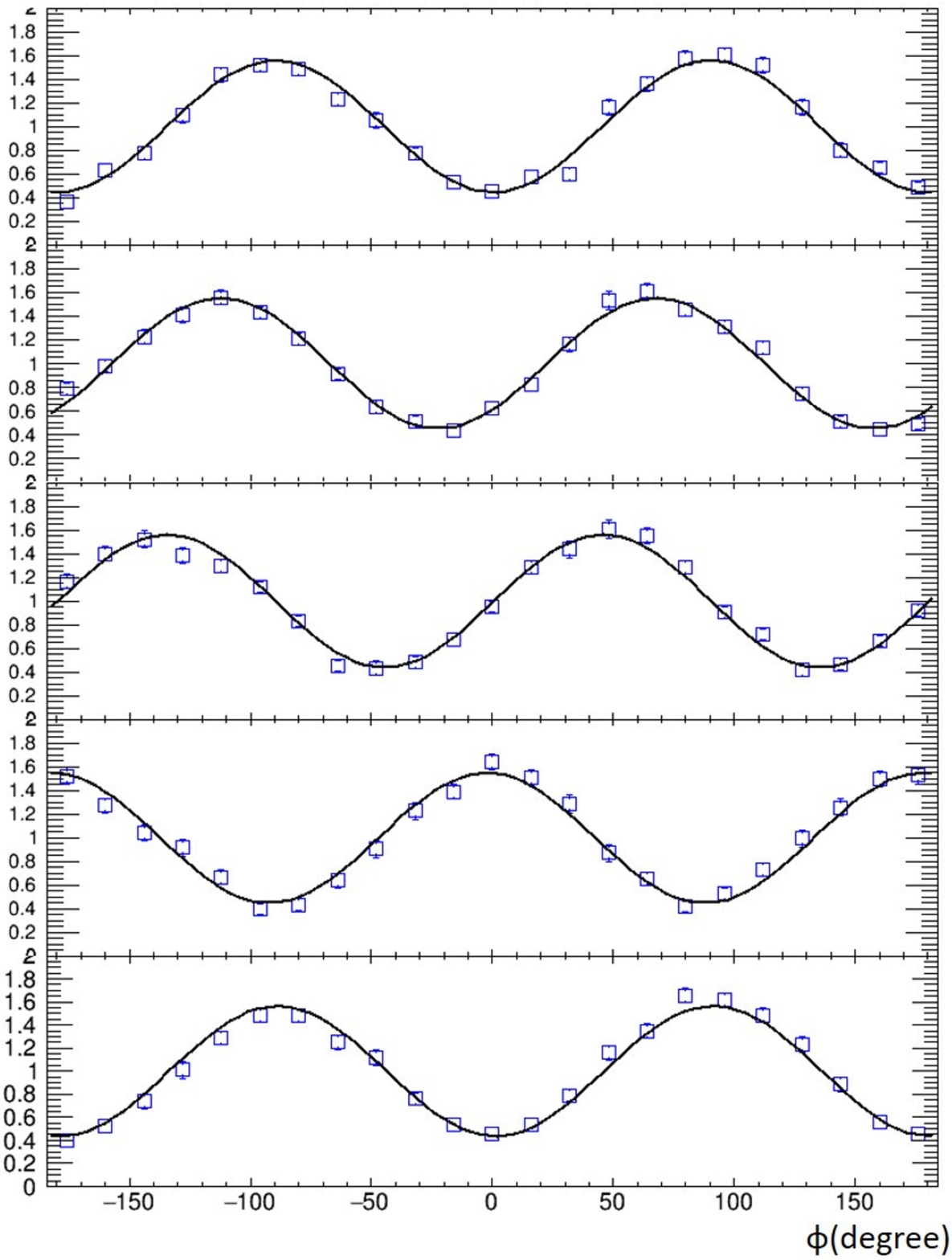}}
  \includegraphics[clip, trim={0 0 {.0\wd0} {.0\ht0}}, width=0.6\textwidth]{SPring-8Exp1CorrectedModulationCurves.eps}
  \endgroup
\caption{
Corrected modulation curves (open squares) and the best fit curves (solid lines). 
The polarization direction of the X-ray beam is 0$^\circ$, $-$22.5$^\circ$, $-$45$^\circ$, $-$90$^\circ$, and $-$180$^\circ$, respectively, from top to bottom.
\label{fig:SPring-8Exp1CorrectedModulationCurves}}
\end{figure}

To obtain the modulation factor, we selected valid events near an energy peak of 134 keV within the FWHM of the energy resolution of the ETCC (29 keV FWHM at 134 keV). The degree of polarization of the incident X-rays is estimated to be 96\% using the theoretical calculation.
Figure \ref{fig:SPring-8Exp1ScatteredGammaDistribution} shows the measured angular distribution of the scattered X-rays, $D^{mes}_{pol}(\cos{\theta}, \phi)$, for a polarization direction of 0$^\circ$.
The calculated azimuthal angle distribution, $N^{mes}_{pol}(\phi)$, from $D^{mes}_{pol}(\cos{\theta}, \phi)$ is plotted in Figure \ref{fig:SPring-8Exp1AzimuthalDistribution}, 
where the simulated azimuthal angle distribution, $N^{sim}_{pol}(\phi)$, reproduces $N^{mes}_{pol}(\phi)$ within approximately 8\%. 
To cancel out the effect due to the non-uniformity of the detector response, we simulated the azimuthal angle distribution for non-polarized photons, $N^{sim}_{unpol}(\phi)$.
Figure \ref{fig:SPring-8Exp1CorrectedModulationCurves} presents the final azimuthal angle distribution corrected by $N^{sim}_{unpol}(\phi)$, $N^{mes}_{cor}(\phi)$ ($=N^{mes}_{pol}(\phi) / N^{sim}_{unpol}(\phi)$), and their best fitting results according to Equation (\ref{eq:modulationcurve}) for five different polarization directions of the X-ray beam.
The obtained modulation factors and polarization angles from the fitting parameters are summarized in Table \ref{tbl:FitResults}. 
The ETCC clearly determined the polarization angles for all the measurements within an accuracy of 1$^\circ$, which is consistent with the polarization directions of the X-ray beam considering the rotation angle accuracy of approximately 0.7$^\circ$.
From these results, we conclude that the modulation factor of the ETCC is in the range of 0.57--0.59 within an error of 0.02.
The ideal value of the modulation factors can be obtained by fitting $N^{sim}_{cor}(\phi)$ ($=N^{sim}_{pol}(\phi) / N^{sim}_{unpol}(\phi)$) and are also included in Table \ref{tbl:FitResults}. 
The differences between the measured and simulated modulation factors are larger than the margin of errors due to the small differences in the azimuthal angle distributions and because the uncertainty in the simulation does not take into account the position resolution of the Compton interaction point in the TPC.

\clearpage

\begin{deluxetable}{cccc}
\tablecaption{Fit results of the polarization parameters for five different polarization directions \label{tbl:FitResults}}
\tablehead{
\colhead{Polarization direction} & \colhead{Polarization angle} & \colhead{Modulation factor} & \colhead{Modulation factor} \\
\colhead{experimental setup} & \colhead{measured} & \colhead{measured} & \colhead{simulated} \\
\colhead{(degree)} & \colhead{(degree)} & \colhead{} & \colhead{} 
}
\startdata
0          &     0.4 $\pm$ 0.9     &   0.58 $\pm$ 0.02 & 0.63 $\pm$ 0.01\\
$-$22.5 &  $-$22.3 $\pm$ 0.8  &   0.58 $\pm$ 0.02 & 0.63 $\pm$ 0.01\\
$-$45   &  $-$44.5 $\pm$ 0.7  &  0.58 $\pm$ 0.02  & 0.62 $\pm$ 0.01\\
$-$90   &  $-$92.2 $\pm$ 1.0  &   0.57 $\pm$ 0.02 & 0.60 $\pm$ 0.01\\
$-$180 & $-$178.7 $\pm$ 0.9  &   0.59 $\pm$ 0.03 & 0.61 $\pm$ 0.01\\
\enddata
\tablecomments{The measured modulation factor is obtained by fitting  $N^{mes}_{cor}(\phi)$, and the simulated modulation factor is obtained by fitting  $N^{sim}_{cor}(\phi)$. These modulation factors and errors are scaled by 0.96, which is the assumed degree of polarization in these measurements.
}
\end{deluxetable}

%\clearpage

\subsection{Polarization measurement for off-axis incidence} \label{sec:experiment2}
\begin{figure}[]
  \centering
  \begingroup
  \sbox0{\includegraphics{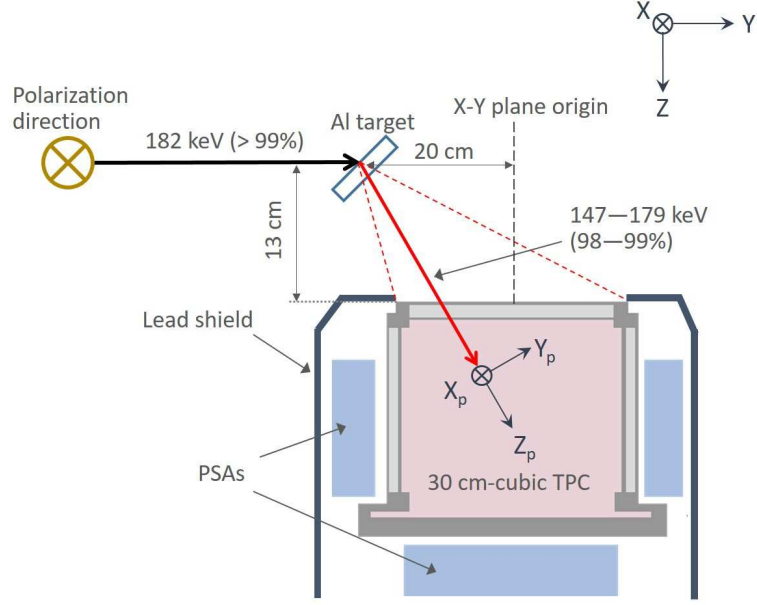}}
  \includegraphics[clip, trim={{.0\wd0} 0 {.0\wd0} {.0\ht0}}, width=0.6\textwidth]{Spring-8Exp2Setup.eps}
  \endgroup
\caption{
A side view in the $Y$--$Z$ plane of the current ETCC setup for the second experiment on BL08W at SPring-8.
The Al target was shifted 20 cm into the upper stream of the beamline than in the previous experiment.
The photon coordinate system $X_{p}Y_{p}Z_{p}$ has the $Z$-axis along the average direction of the incident photons, and the $X$-axis coincides with the $X$--$Y$ plane. 
\label{fig:Spring-8Exp2Setup}}
\end{figure}

\begin{figure}[]
  \centering
  \begingroup
  \sbox0{\includegraphics{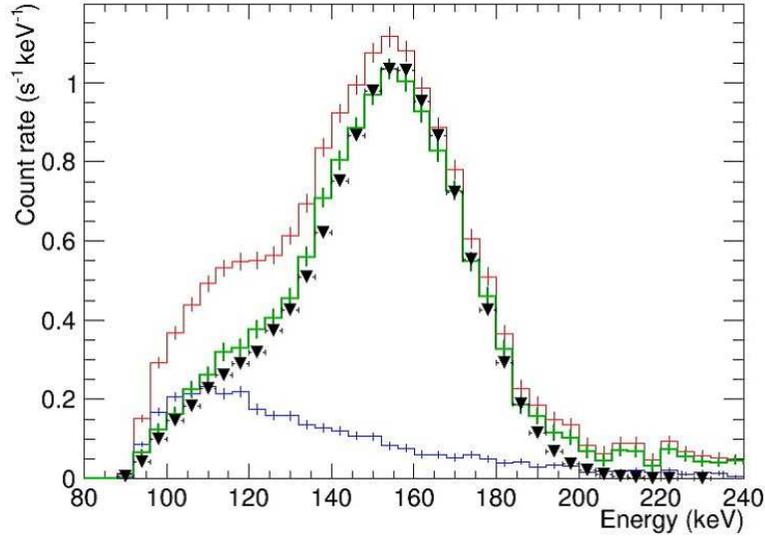}}
  \includegraphics[clip, trim={0 0 {.0\wd0} {.\ht0}}, width=0.6\textwidth]{SPring-8Exp2EnergySpectrum.eps}
  \endgroup
\caption{
Reconstructed energy spectra of the incident photons. The red and blue lines represent the spectra for the Al on-target and off-target data, respectively. 
The green line is the difference between the red and blue lines, which corresponds to the pure energy spectrum of the incident X-rays after subtracting  the noise background due to air scattering. There is good consistency between the green line and the simulated results (the filled triangles) near the energy peak at 154 keV.
\label{fig:SPring-8Exp2EnergySpectrum}}
\end{figure}

\begin{figure}[]
  \centering
  \begingroup
  \sbox0{\includegraphics{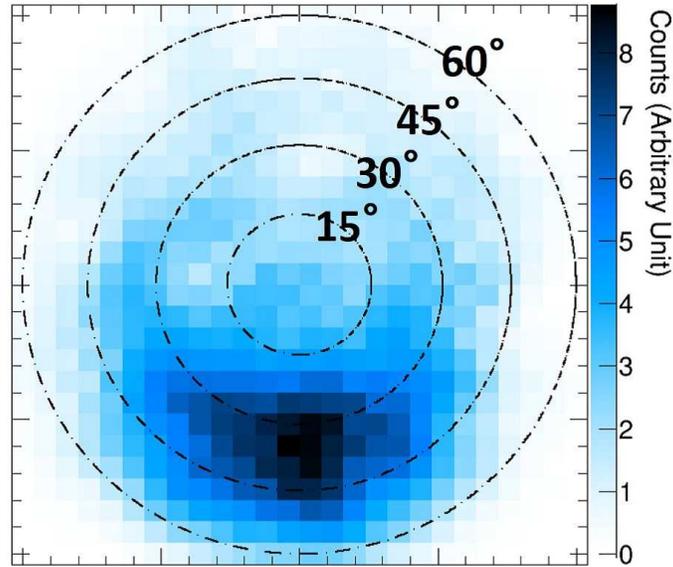}}
  \includegraphics[clip, trim={0 0 {.0\wd0} {.\ht0}}, width=0.5\textwidth]{Spring-8Exp2LambertImage.eps}
  \endgroup
\caption{
Reconstructed image showing the direction of an incident photon. 
The dotted concentric circles represent the incident angle (15$^\circ$, 30$^\circ$, 45$^\circ$, and 60$^\circ$) in the coordinate system of the ETCC. 
The spread in the image corresponds to the spread of the scattered beam. 
An enhancement is seen within the incident angles of 20--40$^\circ$, and most events are concentrated near 30$^\circ$. 
\label{fig:Spring-8Exp2LambertImage}}
\end{figure}

In the next experiment, we measured the modulation factor of the ETCC for an off-axis beam to demonstrate the cancellation of the effect of off-axis incidence.
The experimental setup is shown in Figure \ref{fig:Spring-8Exp2Setup} where the coordinate system of the ETCC, $XYZ$, and the coordinate  system of the incident photon, $X_{p}Y_{p}Z_{p}$, is also defined.
The position of the Al target was shifted 20 cm into the upper stream of the beamline compared to that of the previous experiment.
Scattered X-rays at the Al target entered the ETCC with a tilted incident angle from the $Z$-axis of 20--60$^\circ$. 
Even though the energy was widely spread from 147 keV to 179 keV, the degree of polarization was limited to high at 98--99\% because the forward scattering events were dominant. 
Figure \ref{fig:SPring-8Exp2EnergySpectrum} shows the reconstructed energy spectra of Compton events selected using the same selection method as in the on-axis experiments.
An obvious energy peak near 155 keV appeared after the background subtraction, and the shape of the distribution is well reproduced by the simulation within an error of 10\%. 
An energy peak near 155 keV corresponds to the energy of photons with incident angles of approximately 30$^\circ$, which is consistent with the direction of the reconstructed incident photons projected onto the sphere in Figure \ref{fig:Spring-8Exp2LambertImage}, where the spread of the image is due to the spread of the beam.

\begin{figure}[h]
  \centering
  \begingroup
  \sbox0{\includegraphics{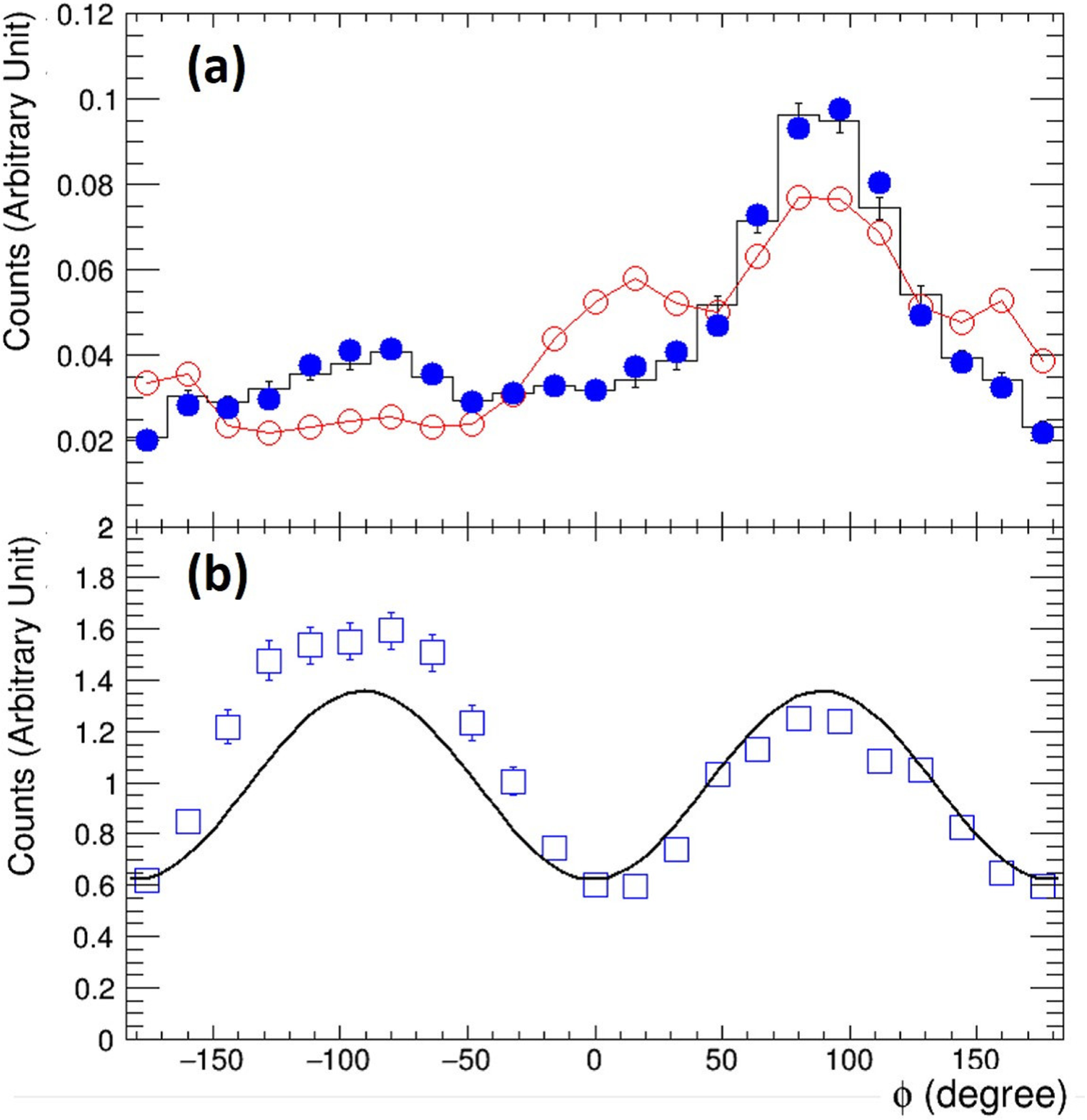}}
  \includegraphics[clip, trim={0 0 {.0\wd0} {.0\ht0}}, width=0.6\textwidth]{SPring-8Exp2AzimuthalDistribution.eps}
  \endgroup
\caption{
(a) Azimuthal angle distributions in the $XYZ$ coordinate system: $N^{mes}_{pol}(\phi)$ (solid line histogram), $N^{sim}_{pol}(\phi)$ (filled circles) and $N^{sim}_{unpol}(\phi)$ (open circles).
(b) Corrected azimuthal angle distribution (open squares) calculated by $N^{mes}_{pol}(\phi)/N^{sim}_{unpol}(\phi)$ and the best fitted curve (solid line). 
\label{fig:SPring-8Exp2AzimuthalDistribution}}
\end{figure}

\begin{figure}[h]
  \centering
  \begingroup
  \sbox0{\includegraphics{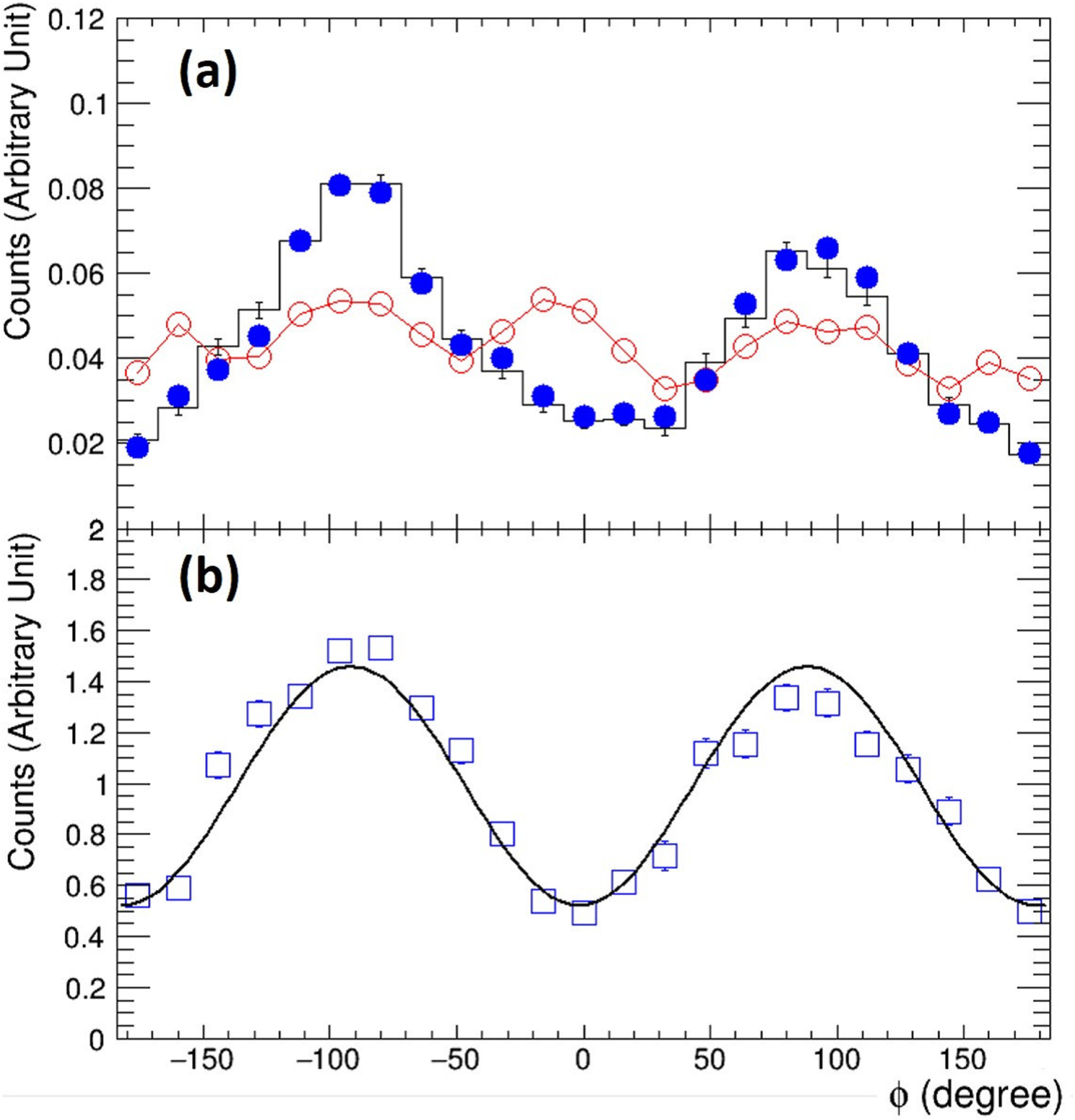}}
  \includegraphics[clip, trim={0 0 {.0\wd0} {.0\ht0}}, width=0.6\textwidth]{SPring-8Exp2CorrectedModulationCurve.eps}
  \endgroup
\caption{
Same plots as shown in Figure \ref{fig:SPring-8Exp2AzimuthalDistribution} in the $X_{p}Y_{p}Z_{p}$ coordinate system. 
\label{fig:SPring-8Exp2CorrectedModulationCurve}}
\end{figure}

\begin{figure}[h]
  \centering
  \begingroup
  \sbox0{\includegraphics{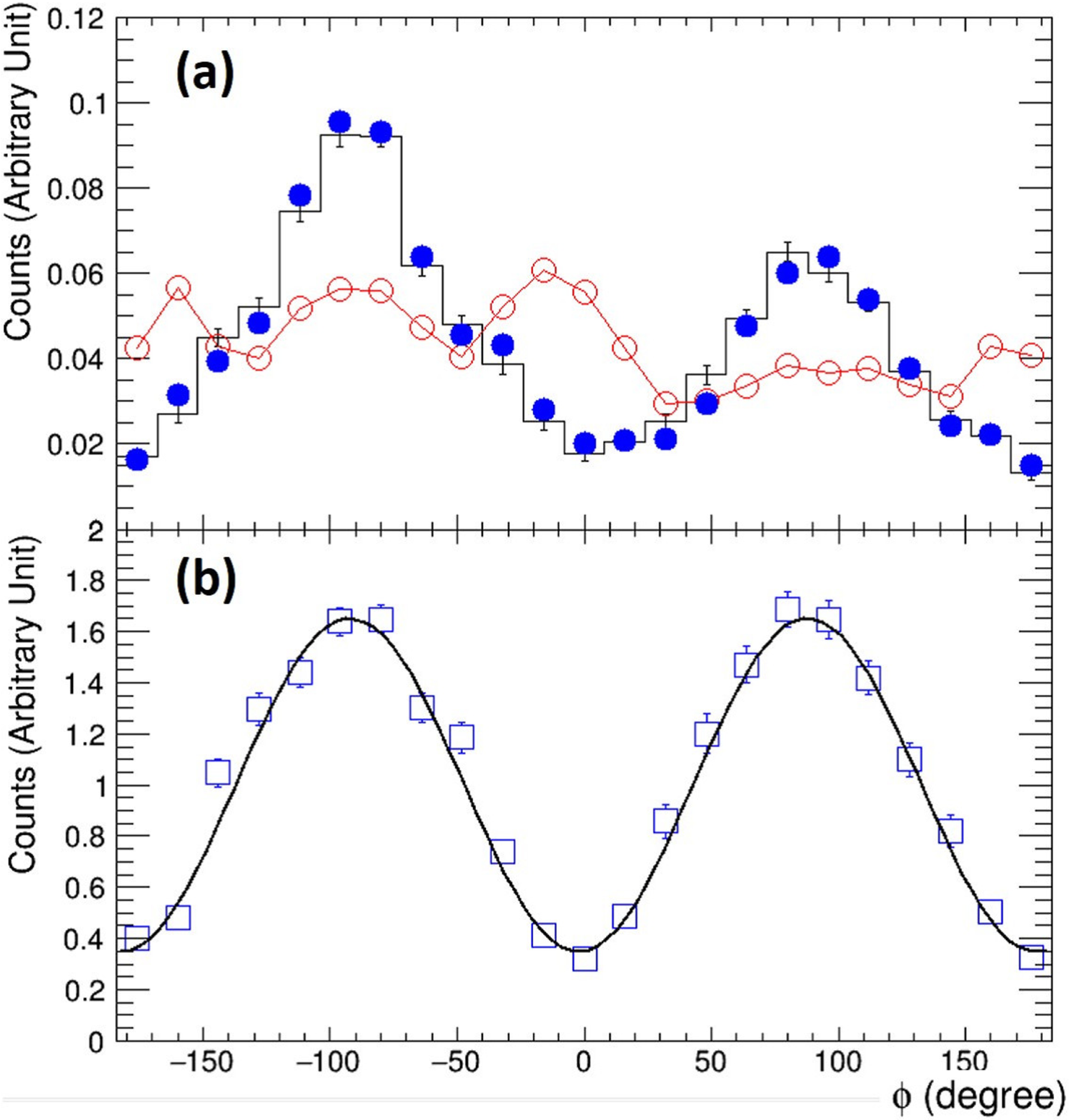}}
  \includegraphics[clip, trim={0 0 {.0\wd0} {.0\ht0}}, width=0.6\textwidth]{SPring-8Exp2CorrectedModulationCurve_0.7.eps}
  \endgroup
\caption{
Same plots as shown in Figure \ref{fig:SPring-8Exp2AzimuthalDistribution} in the $X_{p}Y_{p}Z_{p}$ coordinate system for a $(\cos{\theta})_{max}$ of 0.7.
\label{fig:SPring-8Exp2CorrectedModulationCurve_0.7}}
\end{figure}

To obtain the modulation factor, we selected valid events near the energy peak of 155 keV considering the energy resolution of the ETCC (35 keV FWHM at 155 keV), where the degree of polarization of the incident X-rays is approximately 98\%.
In the next two figures, to clearly show the effect of off-axis incidence, we use the range of 0--1 for the integration over $\cos{\theta}$ in Equation (\ref{eq:Nphi_cos}) because the forward scattering events with small $\theta$ are concentrated in the positive $Y$ direction ($\phi \sim$ 90$^\circ$) and generate large systematic modulations.
Figure \ref{fig:SPring-8Exp2AzimuthalDistribution}(a) shows the measured azimuthal angle distribution in the $XYZ$ coordinate system, $N^{mes}_{pol}(\phi)$, which is distorted by the effect of off-axis incidence. 
If we applied the cancellation of the effect due to the non-uniformity of the detector response to $N^{mes}_{pol}(\phi)$ according to Equation (\ref{eq:efficiency_correction}), the obtained azimuthal angle distribution is far from a symmetrical distribution, as shown in Figure \ref{fig:SPring-8Exp2AzimuthalDistribution}(b), even though the simulated azimuthal angle distribution $N^{sim}_{pol}(\phi)$ reproduces $N^{mes}_{pol}(\phi)$ within 8\%. 
From this response-corrected azimuthal angle distribution, we obtained a modulation factor of 0.33 $\pm$ 0.01.
As mentioned in Section \ref{sec:principles}, we first have to calculate the azimuthal angle distribution in the $X_{p}Y_{p}Z_{p}$ coordinate system, $N^{mes}_{pol}(\phi_{p})$, before canceling out the non-uniform response.
Figure \ref{fig:SPring-8Exp2CorrectedModulationCurve}(a) shows the calculated $N^{mes}_{pol}(\phi_{p})$ using the coordinate transformation matrix assuming that the azimuthal and polar angles of the incident photon are 0$^\circ$ and 30$^\circ$, respectively. 
Then, we canceled out the effect due to the non-uniformity of the detector response to obtain the response-corrected azimuthal angle distribution, $N^{mes}_{cor}(\phi_{p})$, as shown in Figure \ref{fig:SPring-8Exp2CorrectedModulationCurve}(b), where the symmetries are obviously recovered.
By fitting $N^{mes}_{cor}(\phi_{p})$ with Equation (\ref{eq:modulationcurve}), the modulation factor is found to be 0.44 $\pm$ 0.01, which is improved by a factor of 1.3 compared to Figure \ref{fig:SPring-8Exp2AzimuthalDistribution}(b).
Figure \ref{fig:SPring-8Exp2CorrectedModulationCurve_0.7} shows the best results using an appropriate integration range for $\cos{\theta_{p}}$, 0.7 as $(\cos{\theta})_{max}$ in Equation (\ref{eq:Nphi_cos}), and a better modulation factor of 0.65 $\pm$ 0.01 is obtained.

\clearpage

\section{Summary and Discussion} \label{sec:discussion}
\begin{figure}[]
  \centering
  \begingroup
  \sbox0{\includegraphics{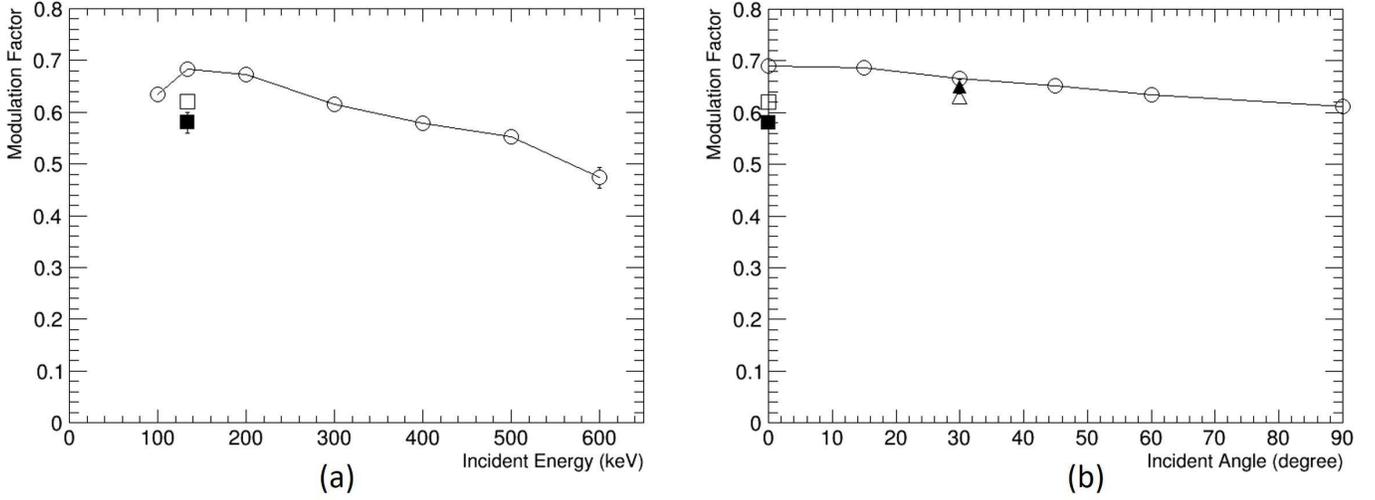}}
  \includegraphics[clip, trim={0 0 {.0\wd0} {.0\ht0}}, width=1.0\textwidth]{SimulatedModulationFactor.eps}
  \endgroup
\caption{
(a) The energy dependence of the simulated modulation factor (circles) of the current ETCC for the on-axis incidence of the parallel incident gamma-rays. The measured (filled squares) and simulated (open squares) modulation factors obtained in Section \ref{sec:experiment1} are also plotted. The modulation factors of the experiment are 9\%--15\% smaller than that of the simulated one at 134 keV due to the non-parallel incidence and the residual backgrounds in the experiment.
(b) Dependence of the simulated modulation factor (circles) on the incident angle for an incident energy of 154 keV. 
It is assumed that the incident direction is defined in the $Y$-$Z$ plane of the ETCC's coordinate system and the polarization direction is in the $X$-$Y$ plane. The modulation factors obtained in Section \ref{sec:experiment1} are plotted at an incident angle of 0$^\circ$ using the same symbols as in panel (a). The measured (filled triangles) and simulated (open triangles) modulation factors obtained in Section \ref{sec:experiment2} are also plotted at an incident angle of 30$^\circ$.
}
\label{fig:SimulatedModulationFactor}
\end{figure}

We present a novel approach for a gamma-ray imaging spectroscopic polarimeter for all-sky surveys using an ETCC, which can perform highly-sensitive polarimetry and spectroscopy for each object, including both persistent and transient objects within its wide FoV of up to 2$\pi$ sr all at once.
The ETCC provides robust solutions to two major difficulties with wide FoV polarimetry, i.e., huge backgrounds coming from all directions and the degradation of the modulation factor due to the effect of off-axis incidence.
It has already been demonstrated that an ETCC can efficiently reject both photon and non-photon backgrounds even in intense radiation conditions similar to space using its excellent imaging performance based on a well-defined PSF and its particle identification using dE/dx \citep{2015ApJ...810...28T}.
Furthermore, the ETCC is expected to maintain its modulation factor without degradation over its wide FoV because it measures all the required information to analytically correct for the effect of off-axis incidence, such as the three-dimensional direction of the scattered photons and the arrival direction and energy of incident photons for each event.
To examine these capabilities, we performed a beam test of the current ETCC using a linearly polarized hard X-ray beam at SPring-8.
Even though there were huge backgrounds of more than twice the polarized X-ray signal, after background rejection, we obtained a modulation factor of 0.58 $\pm$ 0.02 at 134 keV for the quasi-on-axis incidence, which includes the polarized X-ray signal with oblique incident angles of at most 21$^\circ$.
As the greatest impact of this work, we demonstrated for the first time a precise polarization measurement for off-axis incidence with an incident angle of 30$^\circ$ on average; 
we confirmed that the ETCC can correct the distortion of the measured polarization modulation due to off-axis incidence using the measured gamma-ray image, and the obtained modulation factor was 0.65 $\pm$ 0.01 at 154 keV, which is not degraded compared to that of quasi-on-axis incidence.
According to the simulated modulation factors for parallel incident gamma-rays as shown in Figure \ref{fig:SimulatedModulationFactor}, we found that the modulation factor of the ETCC has a maximum of 0.68 near 150 keV, which is the typical photon energy of GRBs, and the modulation factor at 150 keV decreased by only 10\% from 0.68 to 0.62 for an incident angle of 90$^\circ$.
These simulated modulation factors are consistent with the experimental ones even though they are affected by non-parallel incidence and large backgrounds.
Therefore, we conclude that the ETCC can perform wide FoV polarimetry maintaining its high modulation factor of over 0.6 near 150 keV, at least for incident angles less than 30$^\circ$.

Our plan is to perform an all-sky imaging survey using the improved ETCCs in long-duration balloons and satellite experiments with 10 times and 100 times, respectively, better sensitivity than that of COMPTEL \citep{2015ApJ...810...28T}.
In addition, these ETCCs will simultaneously provide polarization measurements of bright objects. 
We have calculated the MDPs in the energy range of 100--300 keV using the performance of the ETCC for balloon observations, such as the effective area of 11 cm$^2$ and the PSF of 23$^\circ$ at an incident energy of 200 keV \citep{Tanimori2017}.
The background rate at balloon altitude is estimated to be approximately 0.11 ph cm$^{-2}$ s$^{-1}$, which includes extragalactic diffuse gamma-rays, atmospheric gamma-rays, and intrinsic gamma-rays calculated by the Geant4 simulations based on results of previous balloon experiments using a small ETCC \citep{2011ApJ...733...13T}. 
The MDPs at 99\% confidence level for the Crab nebula and Cygnus X-1 are calculated as approximately 20\% and 30\%, respectively, according to Equation (\ref{eq:mdp}) in one-day balloon flights with 10 hours of observation. 
Therefore, the ETCC could confirm the observations of {\it INTEGRAL}, which reported that the degree of polarization of the Crab nebula is approximately 40\% \citep{2008Sci...321.1183D, 2008ApJ...688L..29F}.

Thanks to the large FoV of the ETCC, we expect to survey transient objects, in particular typical GRBs with moderate brightness.
Note that, in observations of transients using ETCC, we do not always need simultaneous observations by other satellites to know the energy and direction of the targets, therefore ETCC can perform polarization measurements of all the GRBs in its FoV.
In long-duration balloon experiments, we will use four ETCCs whose effective area would reach 44 cm$^{2}$ (4$\times$11 cm$^{2}$) at an incident energy of 200 keV.
The MDP of an ETCC for GRBs with an intensity 10$^{-5}$ erg cm$^{-2}$ is calculated to be approximately 25\%.
Even if such GRBs had a long duration of several tens of seconds, it is expected that the MDP will degrade less than 5\% thanks to the powerful ETCC background suppression.
If we assume that the degree of polarization of the GRBs is greater than 30\%, we expect to observe approximately 2--3 GRBs during a one-month balloon-flight; this is estimated from the fluence and the duration parameter T$_{90}$ of GRB samples based on the BATSE Current Gamma-Ray Burst Catalog.

\begin{figure}[]
  \centering
  \begingroup
  \sbox0{\includegraphics{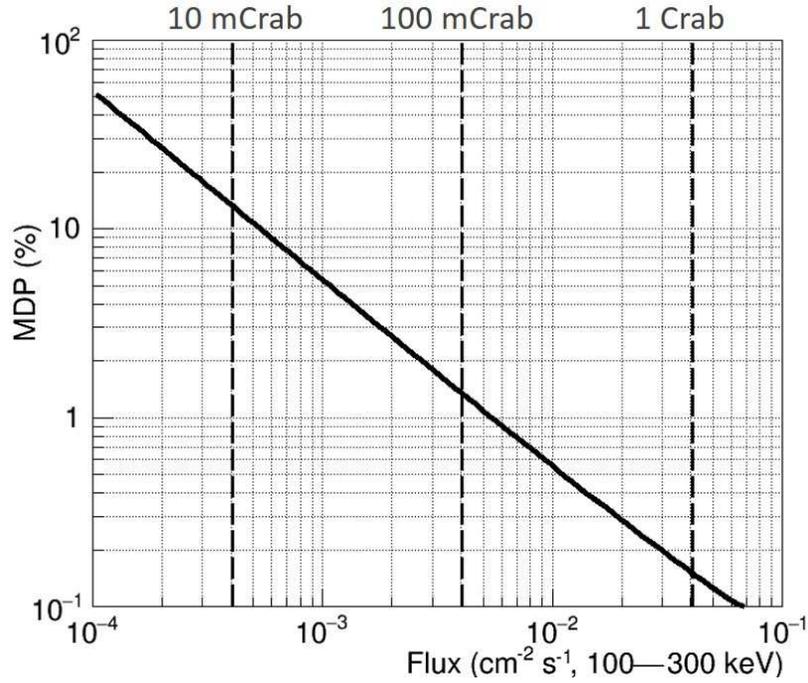}}
  \includegraphics[clip, trim={0 0 {.0\wd0} {.0\ht0}}, width=.6\textwidth]{CalculatedMDPs.eps}
  \endgroup
\caption{
Solid line showing the calculated MDPs of the satellite-ETCC as a function of the source flux in the energy range of 100--300 keV for an observation time of  $10^{7}$ s. The dashed vertical lines represent source fluxes of 1, 1/10, and 1/100 the Crab nebula.
}
\label{fig:CalculatedMDPs}
\end{figure}

For a middle-class satellite experiment, we designed a satellite-ETCC consisting of four 50-cm-cubic ETCCs whose effective area would reach 280 cm$^{2}$ (4$\times$70 cm$^{2}$) at 200 keV \citep{2015ApJ...810...28T}.
Using the effective area of the satellite-ETCC, we calculated the MDPs with respect to the source flux as shown in Figure \ref{fig:CalculatedMDPs}.
We estimate that the MDPs at 13 mCrab would be approximately 10\% for an observation time of $10^{7}$ s, and therefore the satellite-ETCC has the potential to explore polarized serendipity sources as well as major gamma-ray objects, such as AGNs, BBHs, and pulsars.
Simultaneously, we expect that the satellite-ETCC will measure over 20 GRBs that have a fluence of more than $6\times10^{-6}$ erg cm$^{-2}$ and polarization degree of more than 10\% during a one-year observation.
The number of expected GRBs per year is comparable to that of the largest-scale missions, such as POLAR \citep{2011ASTRA...7...43O}; therefore, the satellite-ETCC will contribute to the desired statistical observations of GRB polarizations.

%\clearpage

\acknowledgments
This study was supported by a Grant-in-Aid for Scientific Research from the Ministry of Education, Culture, Sports, Science and Technology (MEXT) of Japan (Grant numbers 21224005, 20244026, 23654067, and 25610042), a Grant-in-Aid from the Global COE program, ``Next Generation Physics, Spun from Universality and Emergence'', from the MEXT of Japan, and a Grant-in-Aid for
JSPS Fellows (Grant number 13J01213). This study was also supported by the ``SENTAN'' program promoted by the Japan Science and Technology Agency (JST). The synchrotron radiation experiments were performed on the BL08W at SPring-8 with the approval of the Japan Synchrotron Radiation Research Institute (JASRI) (Proposal No. 2014B1088). Some of the electronics development was supported by KEK-DTP and the Open-It Consortium. The authors would like to thank Enago (www.enago.jp) for the English language review.

\bibliography{reference}
\allauthors

\end{document}